\documentclass[amsmath, aps, pra, twocolumn, floatfix]{revtex4-2}
\usepackage{graphicx}

\begin{document} 
	
\title{Spatial distributions of the fields in guided normal modes of two coupled parallel optical nanofibers}
	
\author{Fam Le Kien}
\affiliation{Okinawa Institute of Science and Technology Graduate University, Onna, Okinawa 904-0495, Japan}

\author{Lewis Ruks}
\affiliation{Okinawa Institute of Science and Technology Graduate University, Onna, Okinawa 904-0495, Japan}
		
\author{S\'{i}le Nic Chormaic}
\affiliation{Okinawa Institute of Science and Technology Graduate University, Onna, Okinawa 904-0495, Japan}
	
\author{Thomas Busch}
\affiliation{Okinawa Institute of Science and Technology Graduate University, Onna, Okinawa 904-0495, Japan}
	
\date{\today}
	
\begin{abstract}
We study the cross-sectional profiles and spatial distributions of the fields in guided normal modes of two coupled parallel optical nanofibers. We show that the distributions of the components of the field in a guided normal mode of two identical nanofibers are either symmetric or antisymmetric with respect to the radial principal axis and the tangential principal axis in the cross-sectional plane of the fibers. The symmetry of the magnetic field components with respect to the principal axes is opposite to that of the electric field components. We show that, in the case of even $\mathcal{E}_z$-cosine modes, the electric intensity distribution is dominant in the area between the fibers, with a saddle point at the two-fiber center. Meanwhile, in the case of odd $\mathcal{E}_z$-sine modes, the electric intensity distribution at the two-fiber center attains a local minimum of exactly zero. We find that the differences between the results of the coupled mode theory and the exact mode theory are large when the fiber separation distance is small and either the fiber radius is small or the light wavelength is large. We show that, in the case where the two nanofibers are not identical, the intensity distribution is symmetric about the radial principal axis and asymmetric about the tangential principal axis. 
\end{abstract}
	
\maketitle
	
\section{Introduction}

Coupled waveguides form the central working component in numerous optical devices such as multicore fibers, optical directional couplers, polarization splitters, ring resonators, and interferometers \cite{Snyder1983,Marcuse1989, Okamoto2006}. Most of the previous work on coupling between parallel fibers was devoted to conventional fibers where the refractive indices of the core and the cladding differ only slightly from each other and the fiber radius is large compared to the light wavelength \cite{Snyder1983,Marcuse1989, Okamoto2006}. It is desirable to study the properties of guided light fields in coupled subwavelength-diameter optical fibers
due to their increasing relevance in current research efforts \cite{TongNat03}.

Optical nanofibers are tapered fibers that have a subwavelength diameter and significantly differing core and cladding refractive indices \cite{TongNat03}. Such ultrathin fibers allow for a tightly radially confined light field to propagate along the fiber over a long distance (with several millimeters being typical) and to interact efficiently with nearby quantum or classical emitters, absorbers, and scatterers \cite{review2016,review2017,Nayak2018}. Optical nanofibers have been investigated for a variety of applications in nonlinear optics, atomic physics, quantum optics, and nanophotonics \cite{TongNat03,review2016,review2017,Nayak2018}. Nanofibers have been used for trapping of atoms near a nanofiber \cite{onecolor,twocolor,Vetsch2010,Goban2012}, efficient channeling of emission of atoms into guided modes \cite{cesium decay,Nayak2007,Nayak2008}, efficient absorption of guided light by atoms \cite{absorption,Sague2007}, generation of Rydberg states of atoms \cite{Rajasree2020}, and excitation of quadrupole transitions of atoms \cite{quadrupole,Ray2020}. Additionally, slot nanofibers, where the center of the nanofiber has been removed to create two parallel waveguide channels, have been proposed as atom traps \cite{slot}.

Recently, miniaturized optical devices comprising of two twisted or knotted nanofibers have been produced \cite{Glorieux2019}. Coupling between two nanofibers has been studied by using the linear coupling theory \cite{Glorieux2019,CMT}, which is an approximate theory \cite{Snyder1983,Marcuse1989,Okamoto2006}. It has been shown that butt coupling and self coupling \cite{Snyder1983,Marcuse1989,Okamoto2006} could be quite substantial for nanofibers due to the significant mode spread and overlap \cite{CMT}. 

The exact guided normal modes of two coupled dielectric rods can be calculated by the circular harmonics expansion method \cite{Wijngaard1973}. This method has been extended to the case of multicore fibers \cite{Yamashita1985,Kishi1989,Huang1990}. A vector theory that uses the circular harmonics expansion method and the finite-element method has been developed for two-core fibers with radially inhomogeneous core index profiles \cite{Chang1997a}. The propagation constant and the flux density of the field in a guided normal mode have been calculated \cite{Wijngaard1973,Chang1997a,Huang1989}. It has been shown that the coupled mode theory performs well when the separation between the fibers is large \cite{Wijngaard1973,Chang1997a,Huang1989}, and gives satisfactory results even for touching fibers when the fiber radii are large enough \cite{Huang1989}. The polarization patterns \cite{Chang1997a} and the mode cutoffs \cite{Chang1997b} have been investigated.

In this work, we investigate the spatial distributions of the fields in guided normal modes of two coupled parallel optical nanofibers. We find that the distributions of the components of the fields in guided normal modes of two coupled identical nanofibers are either symmetric or antisymmetric with respect to the principal axes of the cross-sectional plane of the fibers. We reveal that the intensity distributions of the fields in guided normal modes of  
two identical fibers attain a local extremum at the two-fiber center that may be used for atom trapping and guiding. Additionally, we show that the discrepancy between the results of the coupled mode theory and the exact theory is large when the fiber separation distance is small and either the fiber radius is small or the light wavelength is large.

The paper is organized as follows. In section \ref{sec:model} we describe the model of two coupled parallel nanofibers 
and present the basic equations for guided normal modes. Section \ref{sec:num} contains the numerical calculations of the spatial distributions of the fields in the guided normal modes. Our conclusions are given in section \ref{sec:summary}.

\section{Two coupled parallel nanofibers}
\label{sec:model}

\begin{figure}[tbh]
	\begin{center}
		\includegraphics{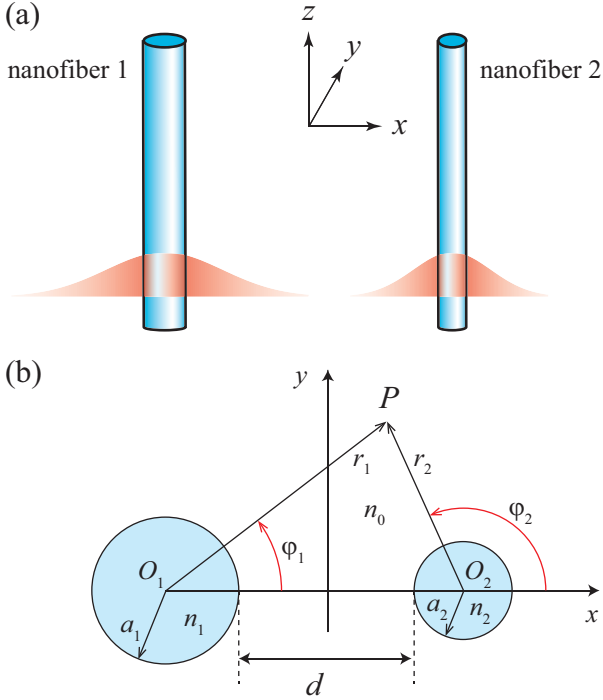}
	\end{center}		
	\caption{(Color online) Two coupled parallel optical nanofibers (a) and the geometry of the system (b). 	 
	}
	\label{fig1}
\end{figure}

We study two vacuum-clad, optical nanofibers that are aligned parallel to each other in the direction of the axis $z$  (see  Fig.~\ref{fig1}). We label the fibers by the indices $j=1,2$. Each nanofiber $j$ is a dielectric cylinder of radius $a_j$ and refractive index $n_j>1$ and is surrounded by an infinite background vacuum or air medium of refractive index $n_0=1$. The diameters of the nanofibers are in the range of hundreds of nanometers. 
An individual nanofiber $j$ can support one or more modes depending on the fiber size parameter $V_j=ka_j\sqrt{n_j^2-n_0^2}$. Here, $k=\omega/c$ is the wave number of light with optical frequency $\omega$ in free space. We are interested in the normal modes of the two-fiber system. We are not interested in the van der Waals interaction between the fibers assuming that they are fixed. 

We introduce the global Cartesian coordinate system $\{x,y,z\}$. Here, the axis $z$ is parallel to the axes $z_1$ and $z_2$ of the fibers, the axis $x$ is perpendicular to the axis $z$ and connects the centers $O_1$ and $O_2$ of the fibers, and the axis $y$ is perpendicular to the axes $x$ and $z$ (see Fig.~\ref{fig1}). The plane $xy$ is the transverse (cross-sectional) plane of the fibers. The axes $x$ and $y$ are called the radial and tangential axes, respectively, of the two-fiber system [see Fig.~\ref{fig1}(b)]. The positions of the fiber centers $O_1$ and $O_2$ on the axis $x$ are $O_1=-(a_1+d_1)$ and $O_2=a_2+d_2$, where $d_1+d_2=d$ is the fiber separation distance. Without loss of generality, we choose $d_1=d_2=d/2$. For each individual fiber $j$, we use the local fiber-based system $\{r_j,\varphi_j\}$ of polar coordinates. 

The normal modes of the coupled fibers are called array modes. We study the array modes of a light field with optical frequency $\omega$, propagating in the $+z$ direction with propagation constant $\beta$. We write the electric and magnetic components of the field as $\mathbf{E}=[\boldsymbol{\mathcal{E}}e^{-i (\omega t-\beta z)}+\mathrm{c.c.}]/2$ and $\mathbf{H}=[\boldsymbol{\mathcal{H}}e^{-i (\omega t-\beta z) t}+\mathrm{c.c.}]/2$, respectively, where $\boldsymbol{\mathcal{E}}$ and $\boldsymbol{\mathcal{H}}$ are the slowly varying complex envelopes. 

The exact theory for the guided normal modes of two parallel dielectric rods has been formulated in \cite{Wijngaard1973}.
The flux density and the beat wavelength for the beating of energy between the guided normal modes have been calculated for the rods with $a_j\gg\lambda/2$ and $n_j/n_0\simeq 1$. We follow the theory of \cite{Wijngaard1973} and use it to treat the spatial distributions of the fields in the guided normal modes of the coupled nanofibers with $a_j\lesssim \lambda/2$ and $n_j/n_0\gtrsim 1.45$.

According to the theory of \cite{Wijngaard1973}, the longitudinal components $\mathcal{E}_z$ and $\mathcal{H}_z$ of the electric and magnetic parts, respectively, of the field in a guided normal mode are given, inside fiber $j=1,2$, as
\begin{eqnarray}\label{g1}
	\mathcal{E}_z&=&\sum_{n=0}^\infty[A_{nj}J_n(h_jr_j)\cos n\varphi_j+E_{nj}J_n(h_jr_j)\sin n\varphi_j],\nonumber\\
	\mathcal{H}_z&=&\sum_{n=0}^\infty[B_{nj}J_n(h_jr_j)\sin n\varphi_j+F_{nj}J_n(h_jr_j)\cos n\varphi_j],\nonumber\\
\end{eqnarray}
and, outside the two fibers, as 
\begin{eqnarray}\label{g2}
	\mathcal{E}_z&=&\sum_{j=1}^2\sum_{n=0}^\infty[C_{nj}K_n(qr_j)\cos n\varphi_j
	\nonumber\\&&\mbox{}
	+G_{nj}K_n(qr_j)\sin n\varphi_j],\nonumber\\
	\mathcal{H}_z&=&\sum_{j=1}^2\sum_{n=0}^\infty[D_{nj}K_n(qr_j)\sin n\varphi_j
	\nonumber\\&&\mbox{}
	+H_{nj}K_n(qr_j)\cos n\varphi_j].
\end{eqnarray}
Here, we have introduced the fiber parameters
\begin{equation}\label{g3}
	h_j=\sqrt{k^2n_j^2-\beta^2},\qquad
	q=\sqrt{\beta^2-k^2n_0^2},
\end{equation}
which determine the scales of the spatial variations of the field inside and outside the fibers.
In Eqs.~(\ref{g1}) and (\ref{g2}), the sets $\{A_{nj},B_{nj},C_{nj},D_{nj}\}$ and $\{E_{nj},F_{nj},G_{nj},H_{nj}\}$ contain the mode expansion coefficients for the $\mathcal{E}_z$-cosine ($x$-polarized) and $\mathcal{E}_z$-sine ($y$-polarized) modes, respectively. The notations $J_n$ and $K_n$ stand for the Bessel functions of the first kind and the modified Bessel functions of the second kind, respectively. 

The transverse components $\mathcal{E}_{x,y}$ and $\mathcal{H}_{x,y}$ of the electric and magnetic parts of the field can be expressed in terms of the longitudinal components $\mathcal{E}_z$ and $\mathcal{H}_z$ as \cite{Snyder1983,Marcuse1989,Okamoto2006}
\begin{eqnarray}\label{g4}
	\mathcal{E}_x&=&\frac{i\beta}{k^2n_{\mathrm{ref}}^2-\beta^2}\left(\frac{\partial}{\partial x}\mathcal{E}_z+\frac{\omega\mu_0}{\beta}\frac{\partial}{\partial y}\mathcal{H}_z\right),\nonumber\\
	\mathcal{E}_y&=&\frac{i\beta}{k^2n_{\mathrm{ref}}^2-\beta^2}\left(\frac{\partial}{\partial y}\mathcal{E}_z-\frac{\omega\mu_0}{\beta}\frac{\partial}{\partial x}\mathcal{H}_z\right),\nonumber\\
	\mathcal{H}_x&=&\frac{i\beta}{k^2n_{\mathrm{ref}}^2-\beta^2}\left(\frac{\partial}{\partial x}\mathcal{H}_z-\frac{\omega\epsilon_0n_{\mathrm{ref}}^2}{\beta}\frac{\partial}{\partial y}\mathcal{E}_z\right),\nonumber\\
	\mathcal{H}_y&=&\frac{i\beta}{k^2n_{\mathrm{ref}}^2-\beta^2}\left(\frac{\partial}{\partial y}\mathcal{H}_z+\frac{\omega\epsilon_0n_{\mathrm{ref}}^2}{\beta}\frac{\partial}{\partial x}\mathcal{E}_z\right).
\end{eqnarray}
Here, $n_{\mathrm{ref}}$ is the spatial distribution of the refractive index in the presence of the two-fiber system,
that is, $n_{\mathrm{ref}}=n_j$ inside fiber $j=1,2$ and $n_{\mathrm{ref}}=n_0$ outside the two fibers.

For the $\mathcal{E}_z$-cosine modes, the expansion coefficients $E_{nj}$, $F_{nj}$, $G_{nj}$, and $H_{nj}$ vanish. For these modes, the coefficients $A_{nj}$ and $B_{nj}$ for the field inside the fibers are given by Eqs.~(\ref{g5}),	
while the coefficients $C_{nj}$ and $D_{nj}$ for the field outside the fibers are nonzero solutions of Eqs.~(\ref{g6}).

For the $\mathcal{E}_z$-sine modes, the expansion coefficients $A_{nj}$, $B_{nj}$, $C_{nj}$, and $D_{nj}$ vanish. For these modes, the coefficients $E_{nj}$ and $F_{nj}$ for the field inside the fibers are given by Eqs.~(\ref{g11}), 	
while the coefficients $G_{nj}$ and $H_{nj}$ for the field outside the fibers are nonzero solutions of Eqs.~(\ref{g12}).

The dispersion equation for the $\mathcal{E}_z$-cosine or $\mathcal{E}_z$-sine modes is $\Delta=0$, where $\Delta$ is the determinant of the system of linear Eqs.~(\ref{g6}) for $C_{nj}$ and $D_{nj}$ or (\ref{g12}) for $G_{nj}$ and $H_{nj}$. The solution to the equation $\Delta=0$ determines the propagation constant $\beta$, which allows us to calculate the other fiber parameters $h_j$ and $q$ [see Eqs.~(\ref{g3})].

Note that the coefficients associated with $C_{nj}$ and $D_{nj}$ in  Eqs.~(\ref{g6}) and with $G_{nj}$ and $H_{nj}$ in Eqs.~(\ref{g12}) are real-valued coefficients. Therefore, 
when we omit a common global phase, we can make
$\{A_{nj},B_{nj},C_{nj},D_{nj}\}$  and, similarly, $\{E_{nj},F_{nj},G_{nj},H_{nj}\}$ to be real-valued coefficients.  Then, the longitudinal field components $\mathcal{E}_z$ and $\mathcal{H}_z$, given by Eqs.~(\ref{g1}) and (\ref{g2}), are real-valued, while the transverse components $(\mathcal{E}_x,\mathcal{E}_y)$ and $(\mathcal{H}_x,\mathcal{H}_y)$, given by Eqs.~(\ref{g4}), are imaginary-valued. Thus, we have
\begin{eqnarray}\label{g13}
	\mathcal{E}_z^*&=&\mathcal{E}_z,\qquad \mathcal{H}_z^*=\mathcal{H}_z,\nonumber\\
	\mathcal{E}_x^*&=&-\mathcal{E}_x,\qquad \mathcal{H}_x^*=-\mathcal{H}_x,\nonumber\\
	\mathcal{E}_y^*&=&-\mathcal{E}_y,\qquad \mathcal{H}_y^*=-\mathcal{H}_y.
\end{eqnarray}
Equations (\ref{g13}) indicate that the longitudinal components $\mathcal{E}_z$ and $\mathcal{H}_z$ of the field in a guided normal mode are $\pi/2$ out of phase with respect to the transverse components $\mathcal{E}_x$, $\mathcal{E}_y$, $\mathcal{H}_x$, and $\mathcal{H}_y$. This relative phase is a typical feature of guided \cite{Snyder1983,Marcuse1989,Okamoto2006} and other transversely confined light fields \cite{Lodahl2017}. 

We consider the particular case where the two fibers are identical, that is, the two fibers have the same radius $a_1=a_2$ and the same core refractive index $n_1=n_2$. In this case, for the $\mathcal{E}_z$-cosine modes, we find
\begin{eqnarray}\label{g14}
	A_{n2}&=& (-1)^n \nu A_{n1}, \qquad B_{n2}=(-1)^n \nu B_{n1},\nonumber\\ 
	C_{n2}&=& (-1)^n \nu C_{n1}, \qquad D_{n2}=(-1)^n \nu D_{n1}, \qquad
\end{eqnarray}  
and, for the $\mathcal{E}_z$-sine modes, we get
\begin{eqnarray}\label{g15}
	E_{n2}&=& (-1)^n \nu E_{n1}, \qquad F_{n2}=(-1)^n \nu F_{n1},\nonumber\\
	G_{n2}&=& (-1)^n \nu G_{n1}, \qquad H_{n2}=(-1)^n \nu H_{n1}, \qquad
\end{eqnarray}  
where $\nu=-1$ or $+1$ corresponds to the even or odd mode, respectively \cite{Wijngaard1973}. 
Then, Eqs.~(\ref{g6}) for the $\mathcal{E}_z$-cosine modes reduce to Eqs.~(\ref{g16}) 
and Eqs.~(\ref{g12}) for the $\mathcal{E}_z$-sine modes lead to Eqs.~(\ref{g17}).

When we perform the transformation $x\to -x$, that is,  $(x,y)\to (-x,y)$, we have $(r_1,\varphi_1)\to(r_2,\pi-\varphi_2)$ and $(r_2,\varphi_2)\to (r_1,\pi-\varphi_1)$. It follows from
the relations (\ref{g14}) and (\ref{g15}) and Eqs.~(\ref{g1}), (\ref{g2}),
and (\ref{g4}) that the field components of the even $\mathcal{E}_z$-cosine and odd $\mathcal{E}_z$-sine modes satisfy the relations \cite{Chang1997a}
\begin{eqnarray}\label{g18}
	\mathcal{E}_x(-x,y)&=&\mathcal{E}_x(x,y), \quad \mathcal{H}_x(-x,y)=-\mathcal{H}_x(x,y),\nonumber\\
	\mathcal{E}_y(-x,y)&=&-\mathcal{E}_y(x,y),\quad
	\mathcal{H}_y(-x,y)=\mathcal{H}_y(x,y),\nonumber\\
	\mathcal{E}_z(-x,y)&=&-\mathcal{E}_z(x,y),\quad
	\mathcal{H}_z(-x,y)=\mathcal{H}_z(x,y),\qquad
\end{eqnarray}
and the field components of the odd $\mathcal{E}_z$-cosine and even $\mathcal{E}_z$-sine modes obey the relations \cite{Chang1997a}
\begin{eqnarray}\label{g19}
	\mathcal{E}_x(-x,y)&=&-\mathcal{E}_x(x,y),\quad \mathcal{H}_x(-x,y)=\mathcal{H}_x(x,y),\nonumber\\
	\mathcal{E}_y(-x,y)&=&\mathcal{E}_y(x,y),\quad
	\mathcal{H}_y(-x,y)=-\mathcal{H}_y(x,y),\nonumber\\
	\mathcal{E}_z(-x,y)&=&\mathcal{E}_z(x,y),\quad
	\mathcal{H}_z(-x,y)=-\mathcal{H}_z(x,y).\qquad
\end{eqnarray}
The symmetry properties of the components of the fields in the guided normal modes of two coupled identical fibers with respect to the transformation $x\to-x$ are summarized in Table \ref{table_x}.

\begin{table}
	\caption{\label{table_x} Symmetry ($+$) and antisymmetry ($-$) of the components of the fields in the guided normal modes of two coupled identical fibers with respect to the transformation $x\to-x$.}
	\begin{ruledtabular}
		\begin{tabular}{lllllll}
			Mode type & $\mathcal{E}_x$ & $\mathcal{E}_y$  &  $\mathcal{E}_z$ & $\mathcal{H}_x$ & $\mathcal{H}_y$  &  $\mathcal{H}_z$ \\
			\hline
			even $\mathcal{E}_z$-cosine	& $+$ & $-$ & $-$ & $-$ & $+$ & $+$ \\
			
			odd $\mathcal{E}_z$-cosine	& $-$ & $+$ & $+$ & $+$ & $-$ & $-$ \\
			
			even $\mathcal{E}_z$-sine	& $-$ & $+$ & $+$ & $+$ & $-$ & $-$ \\
			
			odd $\mathcal{E}_z$-sine	& $+$ & $-$ & $-$ & $-$ & $+$ & $+$ \\
		\end{tabular}
	\end{ruledtabular}
\end{table}

When we perform the transformation $y\to -y$, that is,  $(x,y)\to (x,-y)$, we have $(r_1,\varphi_1)\to(r_1,-\varphi_1)$ and $(r_2,\varphi_2)\to (r_2,-\varphi_2)$. 
It follows from Eqs.~(\ref{g1}), (\ref{g2}), and (\ref{g4}) that
the field components of the $\mathcal{E}_z$-cosine modes satisfy the relations \cite{Chang1997a}
\begin{eqnarray}\label{g20}
	\mathcal{E}_x(x,-y)&=&\mathcal{E}_x(x,y),\quad
	\mathcal{H}_x(x,-y)=-\mathcal{H}_x(x,y),\nonumber\\
	\mathcal{E}_y(x,-y)&=&-\mathcal{E}_y(x,y),\quad
	\mathcal{H}_y(x,-y)=\mathcal{H}_y(x,y),\nonumber\\
	\mathcal{E}_z(x,-y)&=&\mathcal{E}_z(x,y),\quad
	\mathcal{H}_z(x,-y)=-\mathcal{H}_z(x,y),\qquad
\end{eqnarray}
and the field components of the $\mathcal{E}_z$-sine modes obey the relations \cite{Chang1997a}
\begin{eqnarray}\label{g21}
	\mathcal{E}_x(x,-y)&=&-\mathcal{E}_x(x,y),\quad
	\mathcal{H}_x(x,-y)=\mathcal{H}_x(x,y),\nonumber\\
	\mathcal{E}_y(x,-y)&=&\mathcal{E}_y(x,y),\quad
	\mathcal{H}_y(x,-y)=-\mathcal{H}_y(x,y),\nonumber\\
	\mathcal{E}_z(x,-y)&=&-\mathcal{E}_z(x,y),\quad
	\mathcal{H}_z(x,-y)=\mathcal{H}_z(x,y).\qquad
\end{eqnarray}
Note that Eqs.~(\ref{g20}) and (\ref{g21}) remain valid for nonidentical fibers.
The symmetry properties of the components of the fields in the guided normal modes of two coupled fibers with respect to the transformation $y\to-y$ are summarized in Table \ref{table_y}.

\begin{table}
	\caption{\label{table_y} Symmetry ($+$) and antisymmetry ($-$) of the components of the fields in the guided normal modes of two coupled fibers with respect to the transformation $y\to-y$.}
	\begin{ruledtabular}
		\begin{tabular}{lllllll}
			Mode type & $\mathcal{E}_x$ & $\mathcal{E}_y$  &  $\mathcal{E}_z$ & $\mathcal{H}_x$ & $\mathcal{H}_y$  &  $\mathcal{H}_z$ \\
			\hline
			$\mathcal{E}_z$-cosine	& $+$ & $-$ & $+$ & $-$ & $+$ & $-$ \\
			
			$\mathcal{E}_z$-sine	& $-$ & $+$ & $-$ & $+$ & $-$ & $+$  \\
			
		\end{tabular}
	\end{ruledtabular}
\end{table}

Thus, the field components $\mathcal{E}_{x,y,z}$ and $\mathcal{H}_{x,y,z}$ are either symmetric or antisymmetric with respect to the transformations $x\to -x$ and $y\to -y$. This property is a consequence of the fact that the principal axes $x$ and $y$ are the symmetry axes of the  system of two identical fibers. 
We note that the symmetry of the magnetic field components $\mathcal{H}_{x,y,z}$ with respect to the transformation $x\to -x$ or $y\to -y$ is opposite to that of the electric field components $\mathcal{E}_{x,y,z}$. 
The symmetry relations (\ref{g18})--(\ref{g21}) are in agreement with the results of  \cite{Chang1997a}.

It is interesting to note that, in the case of the odd $\mathcal{E}_z$-sine mode, we have the relations  $\mathcal{E}_x(x,y)=-\mathcal{E}_x(x,-y)$, $\mathcal{E}_y(x,y)=-\mathcal{E}_y(-x,y)$, and $\mathcal{E}_z(x,y)=-\mathcal{E}_z(-x,y)$, indicating the antisymmetry of $\mathcal{E}_x$ about the $x$ axis and that of $\mathcal{E}_y$ and $\mathcal{E}_z$ about the $y$ axis. It follows from these relations that, in the case of the odd $\mathcal{E}_z$-sine mode, the electric field at the two-fiber center $(x,y)=(0,0)$ is zero, that is, $\boldsymbol{\mathcal{E}}(0,0)=0$. This feature of the odd $\mathcal{E}_z$-sine mode can be used to produce a local minimum of a blue-detuned optical dipole potential to trap ground-state atoms \cite{Nobel prizers a,Nobel prizers b,Nobel prizers c} or
a local minimum of a ponderomotive optical Rydberg-electron potential
to trap Rydberg atoms \cite{ponderomotive 1,ponderomotive 2}. 
Similarly, we can show that, in the case of  the odd $\mathcal{E}_z$-cosine mode, the magnetic field at the two-fiber center $(x,y)=(0,0)$ is zero, that is, $\boldsymbol{\mathcal{H}}(0,0)=0$.

\section{Numerical calculations}
\label{sec:num}

In this section, we perform numerical calculations for the propagation constants and spatial distributions of the fields in guided normal modes of two parallel vacuum-clad silica-core silica nanofibers. The refractive index of the vacuum cladding is $n_0=1$. The refractive index $n_1=n_2$ of the silica cores of the nanofibers is calculated from the four-term Sellmeier formula for fused silica \cite{Malitson,Ghosh}. In particular, for light with the wavelength $\lambda=800$ nm, we have $n_1=n_2=1.4533$.

According to the previous section, in the case of identical fibers, there are four kinds of normal modes, denoted  
as even $\mathcal{E}_z$-cosine, odd $\mathcal{E}_z$-cosine, even $\mathcal{E}_z$-sine, and odd $\mathcal{E}_z$-sine modes,
or as  $I_0^{(\mathrm{cos})}$, $I_\pi^{(\mathrm{cos})}$, $I_0^{(\mathrm{sin})}$, and $I_\pi^{(\mathrm{sin})}$ modes, respectively \cite{Wijngaard1973}. 
The even $\mathcal{E}_z$-cosine, odd $\mathcal{E}_z$-cosine, even $\mathcal{E}_z$-sine, and odd $\mathcal{E}_z$-sine modes can also be labeled by the letters OO,  OE, EE, and EO, respectively.  These letters indicate the symmetry (E) and antisymmetry (O) of the field component $\mathcal{E}_y$ about the $x$ (first letter) and $y$ (second letter) axes \cite{Chang1997a}. We are interested in the case where the fiber radii are small enough that no more than one normal mode of each of the four kinds can be supported.

\subsection{Propagation constants of guided normal modes of two identical nanofibers}

\begin{figure}[tbh]
	\begin{center}
		\includegraphics{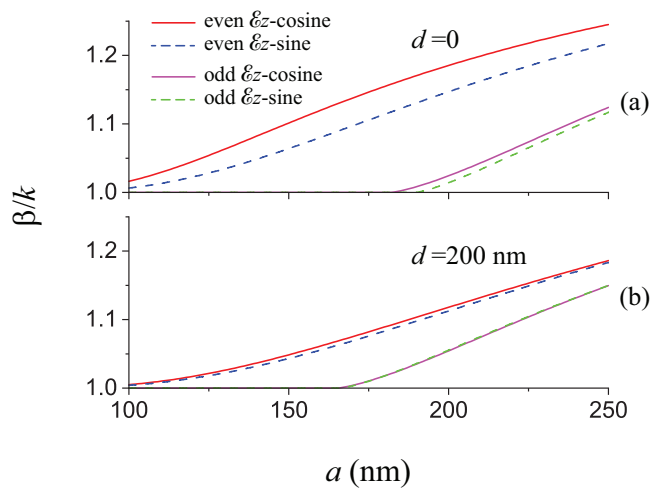}
	\end{center}
	\caption{(Color online) Propagation constant $\beta$, normalized to the free-space wave number $k$, as a function of the fiber radius $a$ in the case of two identical nanofibers. The wavelength of light is $\lambda=800$ nm. The fiber separation distance is $d=0$ (a) and 200 nm (b). The refractive index of the fibers is $n_1=n_2=1.4533$ and of the surrounding medium is $n_0=1$. 
	}
	\label{fig2}
\end{figure}

\begin{figure}[tbh]
	\begin{center}
		\includegraphics{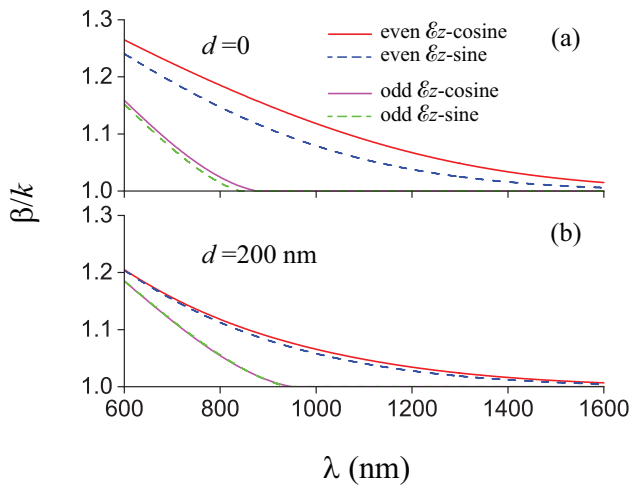}
	\end{center}
	\caption{(Color online) 
		Propagation constant $\beta$, normalized to the free-space wave number $k$, as a function of the wavelength $\lambda$ of light in the case of two identical nanofibers. The fiber radius is $a=200$ nm. 
		The fiber separation distance is $d=0$ (a) and 200 nm (b). 
		The refractive index $n_1=n_2$ of the nanofibers is calculated from the four-term Sellmeier formula for fused silica \cite{Malitson,Ghosh}. The refractive index of the surrounding medium is $n_0=1$.   
	}
	\label{fig3}
\end{figure}

\begin{figure}[tbh]
	\begin{center}
		\includegraphics{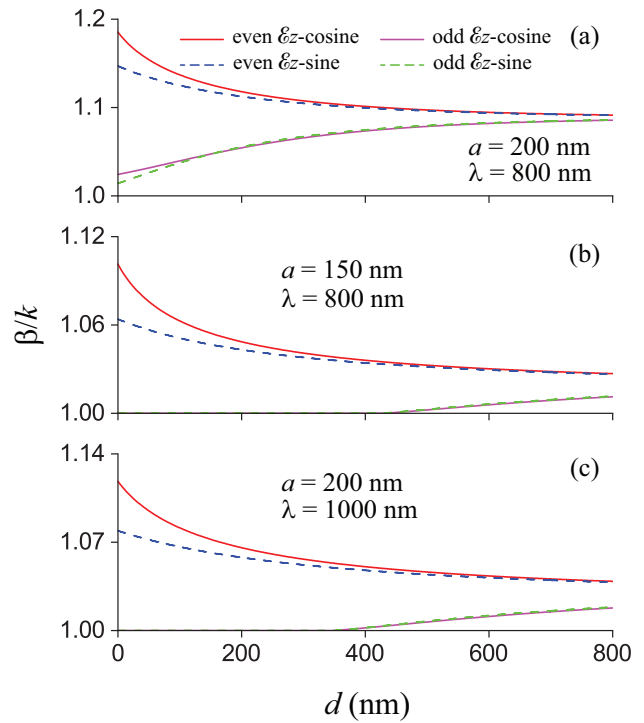}
	\end{center}
	\caption{(Color online) 
		Propagation constant $\beta$, normalized to the free-space wave number $k$, as a function of the fiber separation distance $d$ in the case of two identical nanofibers. 	
		The fiber radius and the light wavelength are (a) $a=200$ nm and $\lambda=800$ nm, (b)  $a=150$ nm and $\lambda=800$ nm, and (c)  $a=200$ nm and $\lambda=1000$ nm.
		Other parameters are as in Figs.~\ref{fig2} and \ref{fig3}. 
	}
	\label{fig4}
\end{figure}

We assume that the two nanofibers have the same fiber radius, that is, $a_1=a_2=a$.
We plot in Figs.~\ref{fig2}--\ref{fig4} the propagation constant $\beta$ as functions of the fiber radius $a$,
the light wavelength $\lambda$, and the fiber separation distance $d$. We observe that there are four guided normal modes, identified as even $\mathcal{E}_z$-cosine, odd $\mathcal{E}_z$-cosine, even $\mathcal{E}_z$-sine, and odd $\mathcal{E}_z$-sine modes \cite{Wijngaard1973}.

We observe from Figs.~\ref{fig2}--\ref{fig4} that there are two pairs of adjacent curves. The upper pair corresponds to the even modes and the lower pair to the odd modes. This indicates that the propagation constant of an even mode (see the upper pair of curves) is larger than that of the corresponding odd mode (see the lower pair of curves). The differences between the propagation constants for the odd $\mathcal{E}_z$-cosine and odd $\mathcal{E}_z$-sine modes (see the lower pair of curves) are smaller than those for the even $\mathcal{E}_z$-cosine and even $\mathcal{E}_z$-sine modes (see the upper pair of curves).  We see from Fig.~\ref{fig4} that the differences between the propagation constants for the four guided array modes reduce with increasing fiber separation distance $d$. We observe from Figs.~\ref{fig2} and \ref{fig3} that the odd $\mathcal{E}_z$-cosine and odd $\mathcal{E}_z$-sine modes have cutoffs but the even $\mathcal{E}_z$-cosine and even $\mathcal{E}_z$-sine modes have no cutoff \cite{Wijngaard1973,Chang1997b}. The reason is the following:

For a single propagation direction $+z$, each single-mode nanofiber can support a superposition of two fundamental modes HE$_{11}$ that are quasilinearly polarized along the $x$ and $y$ axes and are called as $\mathcal{E}_z$-cosine and $\mathcal{E}_z$-sine modes, respectively. 
We expect that two coupled parallel single-mode nanofibers can support up to four guided normal modes. We introduce the notation $\mathbf{e}_j^{(p)}$ for the profile function of the single-fiber mode with the quasilinear polarization $p=x,y$ of nanofiber $j=1,2$. According to the coupled mode theory \cite{Okamoto2006}, there is no coupling between the principal $x$ and $y$ polarizations. For an appropriate choice of the global phases of the mode functions, the approximate profile functions of the guided normal modes can be given as 
$\mathbf{e}_{\pm}^{(p)}\simeq\mathbf{e}_1^{(p)}\pm\mathbf{e}_2^{(p)}$, where the sign $+$ and $-$ corresponds to the even and odd modes, respectively. Note that $\mathbf{e}_1^{(p)}(\mathbf{r})=\mathbf{e}_2^{(p)}(\mathbf{r}+\overrightarrow{O_1O_2})$. 
When the fiber radius is small enough or the light wavelength is large enough, we can use the approximation $\mathbf{e}_1^{(p)}(\mathbf{r})\simeq \mathbf{e}_2^{(p)}(\mathbf{r})$. In these limits, the even array modes with the profile functions $\mathbf{e}_{+}^{(p)}\simeq\mathbf{e}_1^{(p)}+\mathbf{e}_2^{(p)}$ approach the modes of single nanofibers, while the odd array modes with the profile functions $\mathbf{e}_{-}^{(p)}\simeq\mathbf{e}_1^{(p)}-\mathbf{e}_2^{(p)}$ are widely spread out in the outside of the nanofibers. Consequently, the propagation constant of an odd mode is smaller than that of the corresponding even mode (see Figs.~\ref{fig2}--\ref{fig4}).  
When the propagation constant $\beta$ of an odd mode achieves the free space value $k$, the mode is not guided, and a cutoff is observed. We note that the position of the cutoff is determined by the solution to the equation $\beta=k$, where the propagation constant lies on the free-space light line.

Comparison between Figs.~\ref{fig2}(a) and \ref{fig2}(b) and between Figs.~\ref{fig3}(a) and \ref{fig3}(b) shows that the cutoff values of the fiber radius $a$ and the light wavelength $\lambda$ for the odd $\mathcal{E}_z$-cosine and $\mathcal{E}_z$-sine modes depend on the fiber separation distance $d$. A smaller $d$ leads to a larger cutoff value of the fiber radius $a$ and to a smaller cutoff value of the light wavelength $\lambda$. We observe from Fig.~\ref{fig4}(a) that, in the case where the fiber radius $a$ is large enough or, equivalently, the light wavelength is small enough, there is no cutoff of the guided normal modes. However, Figs.~\ref{fig4}(b) and \ref{fig4}(c) show that, in the case where the fiber radius $a$ is small enough or, equivalently, the light wavelength is large enough, a cutoff of an odd guided normal mode may appear at a nonzero fiber separation distance $d$. Comparison between the solid and dashed curves of Figs.~\ref{fig2}--\ref{fig4} shows that the difference between the propagation constants of the $\mathcal{E}_z$-cosine and $\mathcal{E}_z$-sine modes reduces with increasing fiber separation distance $d$. This feature is a consequence of the fact that the difference between the propagation constants of the $x$- and $y$-polarized array modes is determined by the coupling between the nanofibers, which depends on the mode overlap and consequently reduces with increasing $d$.

\subsection{Spatial profiles of the fields in the guided normal modes of two identical nanofibers}

In this subsection, we study the spatial distributions of the fields in the guided normal modes of two identical nanofibers.
We display the cross-sectional profiles of the electric intensity distributions $|\mathcal{E}|^2$ for different guided array modes. We also show the dependencies of the components $\mathcal{E}_x$, $\mathcal{E}_y$, and $\mathcal{E}_z$ 
and the intensity $|\mathcal{E}|^2$ of the electric field on the $x$ and $y$ coordinates (since $\mathcal{E}_x$ and $\mathcal{E}_y$ are imaginary-valued and $\mathcal{E}_z$ is real-valued, we plot $\mathrm{Im}\,(\mathcal{E}_x)=-i\mathcal{E}_x$, $\mathrm{Im}\,(\mathcal{E}_y)=-i\mathcal{E}_y$, and $\mathrm{Re}\,(\mathcal{E}_z)=\mathcal{E}_z$).

\subsubsection{Even $\mathcal{E}_z$-cosine mode}	

\begin{figure}[tbh]
	\begin{center}
		\includegraphics{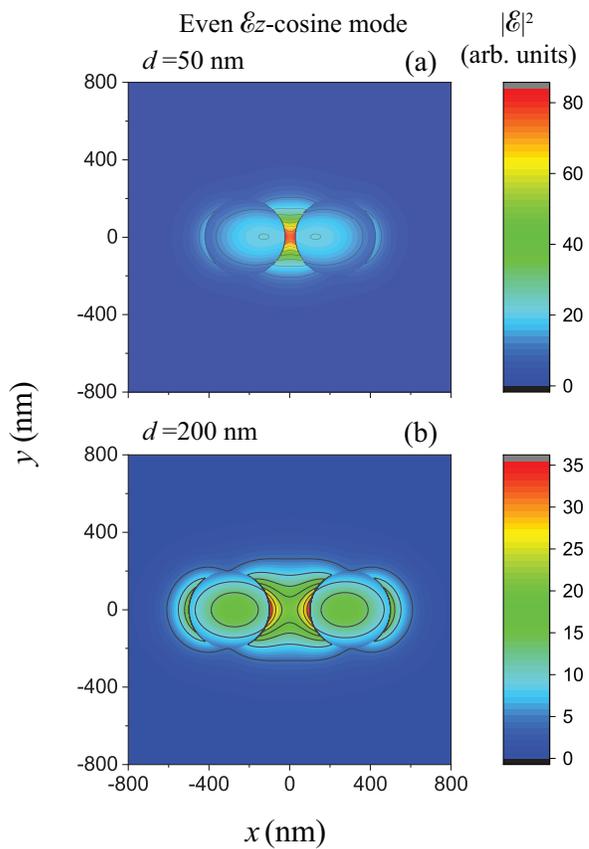}
	\end{center}
	\caption{(Color online) Cross-sectional profile of the electric intensity distribution 
		$|\mathcal{E}|^2$ of the field in the even $\mathcal{E}_z$-cosine mode of two identical parallel nanofibers. The fiber radius is $a=200$ nm. The fiber separation distance is $d=50$ nm (a) and 200 nm (b). The power of light is the same in the two cases. Other parameters are as in Fig.~\ref{fig2}.}
	\label{fig5}
\end{figure}

We plot in Fig.~\ref{fig5} the cross-sectional profile of the electric intensity distribution $|\mathcal{E}|^2$ of the field in the even $\mathcal{E}_z$-cosine mode of two identical parallel nanofibers. We observe from the figure that the intensity distribution $|\mathcal{E}|^2$ is symmetric with respect to the principal axes $x$ and $y$. Figure \ref{fig5} shows that the electric field intensity is dominant in the area between the fibers. This feature can be used to attract atoms \cite{Nobel prizers a,Nobel prizers b,Nobel prizers c} using a single red-detuned  array-mode light field.

\begin{figure}[tbh]
	\begin{center}
		\includegraphics{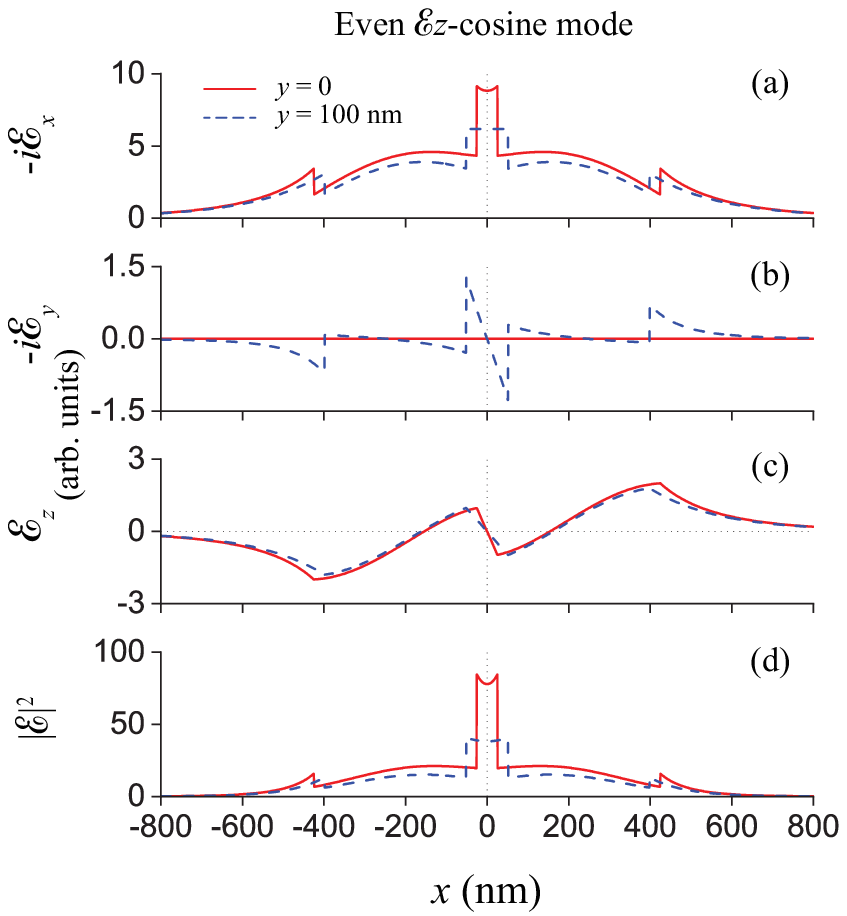}
	\end{center}
	\caption{(Color online) 
		Dependencies of the components $\mathcal{E}_x$, $\mathcal{E}_y$, and $\mathcal{E}_z$ and the intensity $|\mathcal{E}|^2$ of the electric field in the even $\mathcal{E}_z$-cosine mode of two identical fibers on the $x$ coordinate. The fiber radius is $a=200$ nm. The fiber separation distance is $d=50$ nm. The parameters used are the same as for Fig.~\ref{fig5}(a).
	}
	\label{fig6}
\end{figure}

\begin{figure}[tbh]
	\begin{center}
		\includegraphics{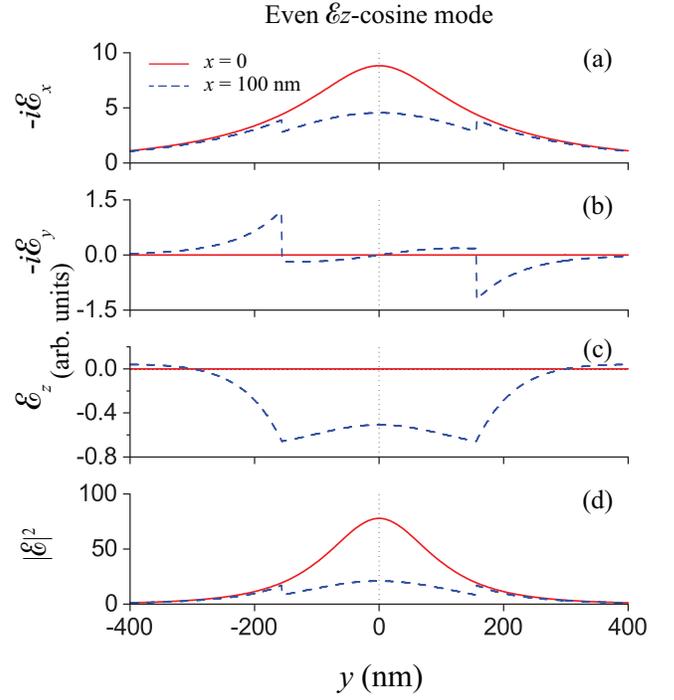}
	\end{center}
	\caption{(Color online) 
		Dependencies of the components $\mathcal{E}_x$, $\mathcal{E}_y$, and $\mathcal{E}_z$ and the intensity $|\mathcal{E}|^2$ of the electric field in the even $\mathcal{E}_z$-cosine mode of two identical fibers on the $y$ coordinate. The fiber radius is $a=200$ nm. The fiber separation distance is $d=50$ nm. The parameters used are the same as for Fig.~\ref{fig5}(a).
	}
	\label{fig7}
\end{figure}

We show in Figs.~\ref{fig6} and \ref{fig7} the dependencies of the components $\mathcal{E}_x$, $\mathcal{E}_y$, and $\mathcal{E}_z$ and the intensity $|\mathcal{E}|^2$ of the electric field in the even $\mathcal{E}_z$-cosine mode on the $x$ and $y$ coordinates. Figure \ref{fig6} shows that $\mathcal{E}_x$ is symmetric and  $\mathcal{E}_y$ and $\mathcal{E}_z$ are antisymmetric with respect to the $x$ coordinate. We observe from Fig.~\ref{fig7} that $\mathcal{E}_x$ and $\mathcal{E}_z$ are symmetric and $\mathcal{E}_y$ is antisymmetric with respect to the $y$ coordinate. Comparison between the scales of the vertical axes of the figures shows that all the three components $\mathcal{E}_x$, $\mathcal{E}_y$, and $\mathcal{E}_z$ of the field are significant, while $\mathcal{E}_x$ [see Figs.~\ref{fig6}(a) and \ref{fig7}(a)] is dominant. These features are in agreement with the fact that, in the case of single fibers, the $\mathcal{E}_z$-cosine modes are quasilinearly polarized along the $x$ axis \cite{Snyder1983,Marcuse1989,Okamoto2006}. 

Figures \ref{fig5}--\ref{fig7} show that a significant portion of the field is in the outside of the nanofibers.
The figures also show that abrupt changes of the fields occur at the surfaces of the fibers. Such dramatic changes are due to the boundary conditions and the sharp contrast between the refractive index $n_1=n_2$ of the silica nanofibers and the refractive index $n_0=1$ of the vacuum medium outside the nanofibers.

The solid curve of Fig.~\ref{fig6}(d) shows that the peaks of the intensity distribution occur at the facing points $(x,y)=(\pm d/2,0)$. Meanwhile, the center $(x,y)=(0,0)$ of the two-fiber system is a saddle point [see Figs.~\ref{fig6}(d) and \ref{fig7}(d)]. Despite these facts, the intensity of the electric field in the area between the two fibers is substantially higher than that in the surrounding area. This behavior of the electric field intensity distribution can, as already mentioned above, be used to attract atoms using red-detuned light to produce an attractive optical dipole potential.

\subsubsection{Odd $\mathcal{E}_z$-cosine mode}

\begin{figure}[tbh]
	\begin{center}
		\includegraphics{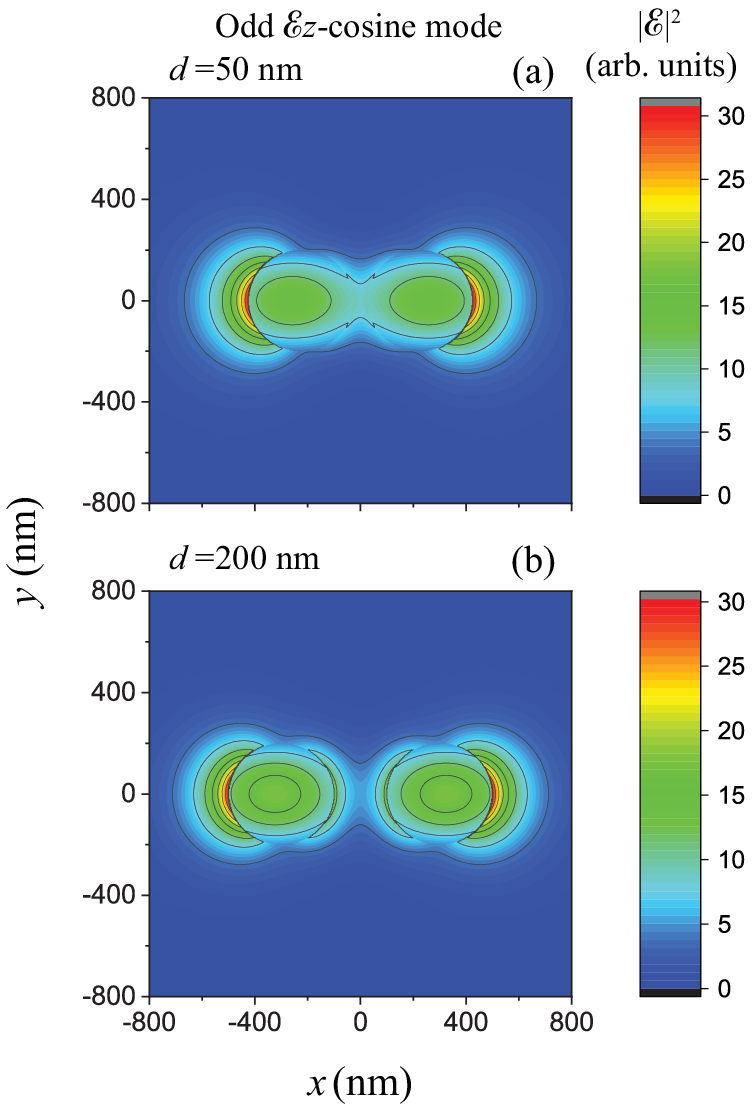}
	\end{center}
	\caption{(Color online) 
		Cross-sectional profile of the electric intensity distribution 
		$|\mathcal{E}|^2$ of the field in the odd $\mathcal{E}_z$-cosine mode of two identical parallel nanofibers. The parameters used are the same as for Fig.~\ref{fig5}.
	}
	\label{fig8}
\end{figure}

We depict in Fig.~\ref{fig8} the cross-sectional profile of the electric intensity distribution 
$|\mathcal{E}|^2$ of the field in the odd $\mathcal{E}_z$-cosine mode of two identical parallel nanofibers. The figure shows that the intensity distribution $|\mathcal{E}|^2$ is symmetric with respect to the principal axes $x$ and $y$. We observe that the electric field intensity is dominant in the outer vicinities of the left-side surface of the left-side fiber and the right-side surface of the right-side fiber.

\begin{figure}[tbh]
	\begin{center}
		\includegraphics{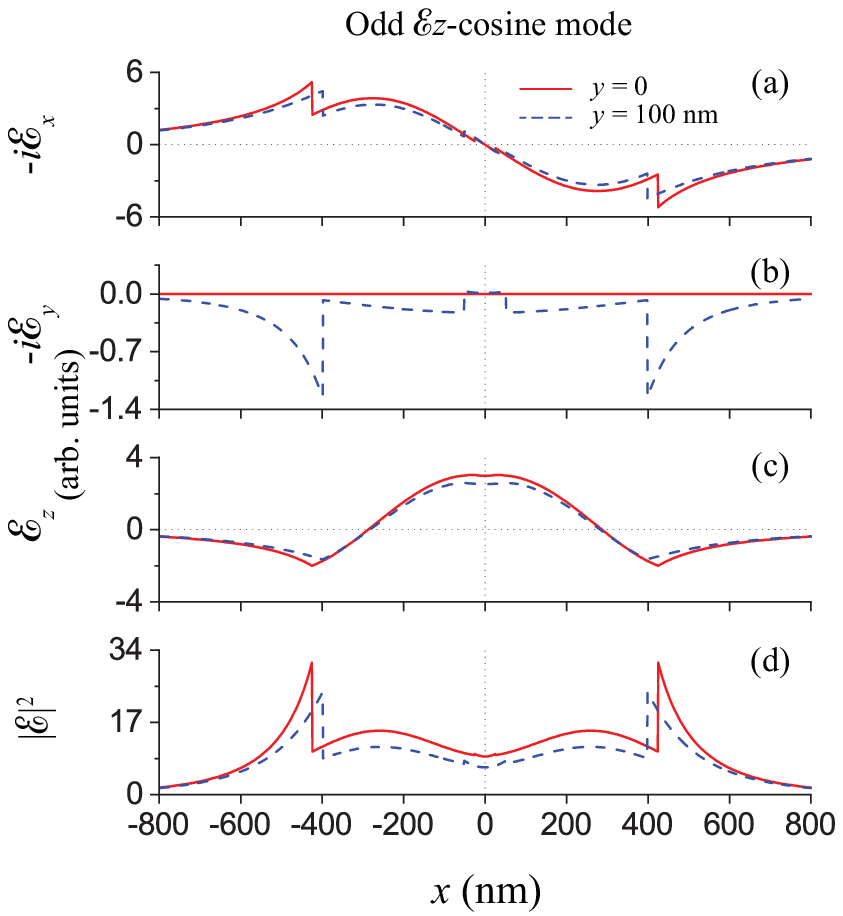}
	\end{center}
	\caption{(Color online) 
		Dependencies of the components $\mathcal{E}_x$, $\mathcal{E}_y$, and $\mathcal{E}_z$ and the intensity $|\mathcal{E}|^2$ of the electric field in the odd $\mathcal{E}_z$-cosine mode of two identical fibers on the $x$ coordinate. The fiber radius is $a=200$ nm. The fiber separation distance is $d=50$ nm. The parameters used are the same as for Fig.~\ref{fig8}(a).
	}
	\label{fig9}
\end{figure}

\begin{figure}[tbh]
	\begin{center}
		\includegraphics{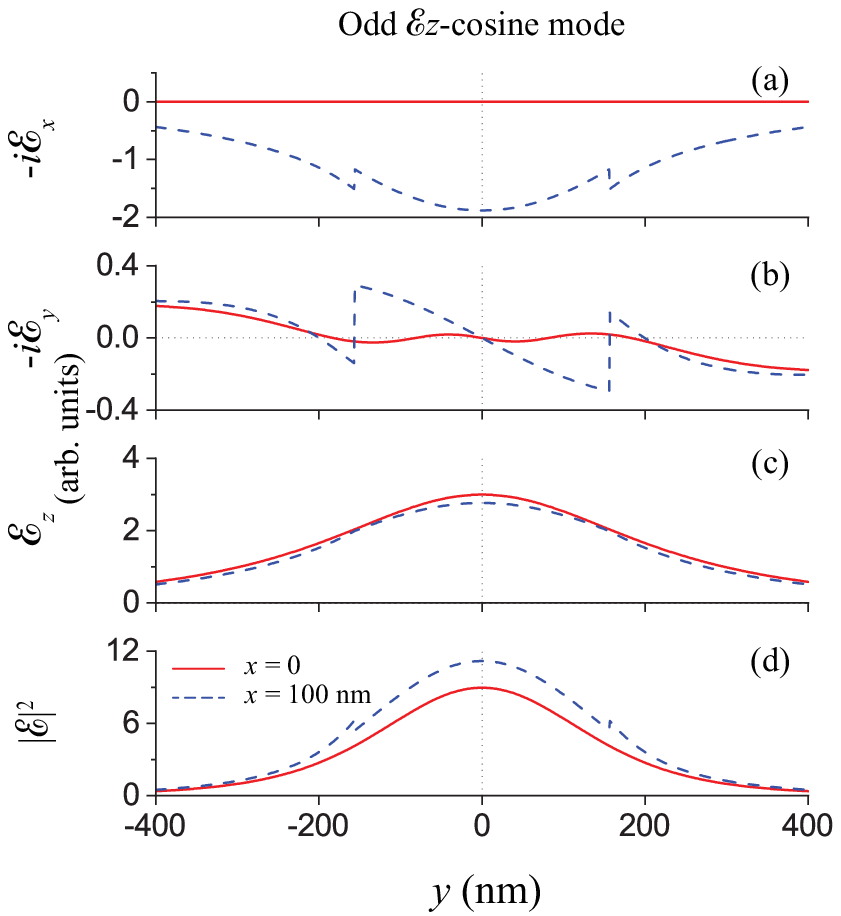}
	\end{center}
	\caption{(Color online) 
		Dependencies of the components $\mathcal{E}_x$, $\mathcal{E}_y$, and $\mathcal{E}_z$ and the intensity $|\mathcal{E}|^2$ of the electric field in the odd $\mathcal{E}_z$-cosine mode of two identical fibers on the $y$ coordinate. The fiber radius is $a=200$ nm. The fiber separation distance is $d=50$ nm. The parameters used are the same as for Fig.~\ref{fig8}(a).
	}
	\label{fig10}
\end{figure}

We display in Figs.~\ref{fig9} and \ref{fig10} the dependencies of the components $\mathcal{E}_x$, $\mathcal{E}_y$, and $\mathcal{E}_z$ and the intensity $|\mathcal{E}|^2$ of the electric field in the odd $\mathcal{E}_z$-cosine mode on the $x$ and $y$ coordinates. Figure \ref{fig9} shows that $\mathcal{E}_x$ is antisymmetric and $\mathcal{E}_y$ and $\mathcal{E}_z$ are symmetric with respect to the coordinate $x$. We observe from Fig.~\ref{fig10} that $\mathcal{E}_x$ and $\mathcal{E}_z$ are symmetric and $\mathcal{E}_y$ is antisymmetric with respect to the coordinate $y$. We see from the scales of the vertical axes in Figs.~\ref{fig9} and \ref{fig10} that all the three components $\mathcal{E}_x$, $\mathcal{E}_y$, and $\mathcal{E}_z$ of the field are significant.

\subsubsection{Even $\mathcal{E}_z$-sine mode}

\begin{figure}[tbh]
	\begin{center}
		\includegraphics{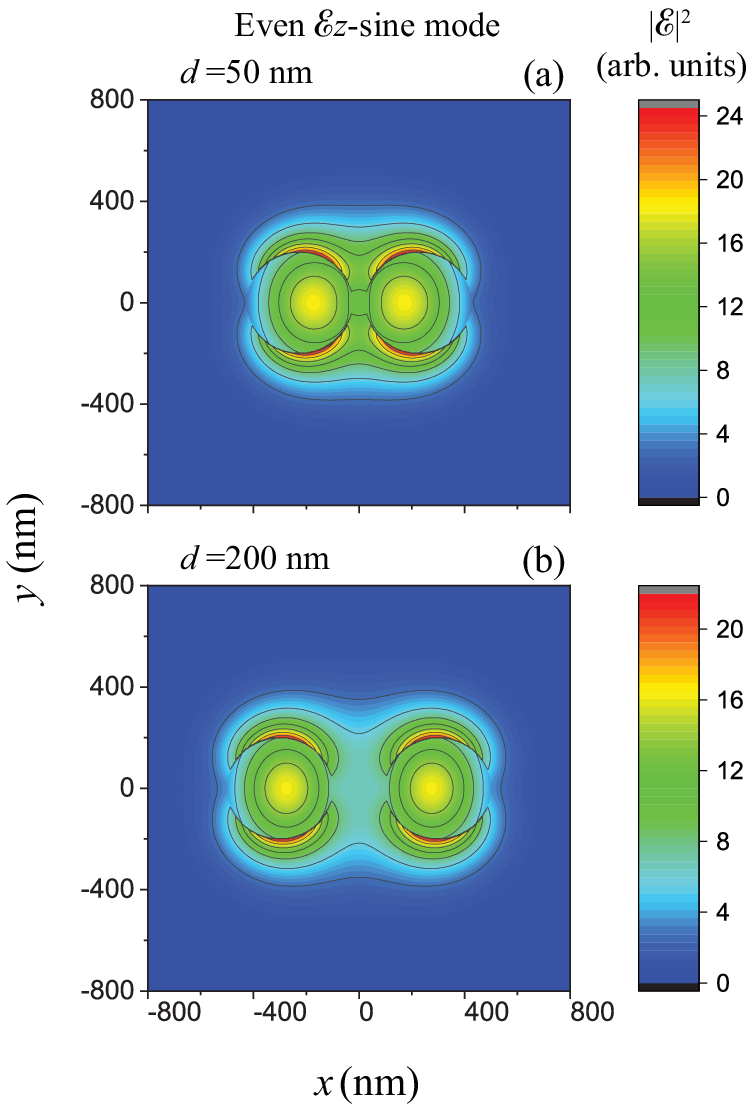}
	\end{center}
	\caption{(Color online) 
		Cross-sectional profile of the electric intensity distribution 
		$|\mathcal{E}|^2$ of the field in the even $\mathcal{E}_z$-sine mode of two identical parallel nanofibers. The parameters used are the same as for Fig.~\ref{fig5}.
	}
	\label{fig11}
\end{figure}

We show in Fig.~\ref{fig11} the cross-sectional profile of the electric intensity distribution 
$|\mathcal{E}|^2$ of the field in the even $\mathcal{E}_z$-sine mode of two identical parallel nanofibers. It is clear from the figure that the intensity distribution $|\mathcal{E}|^2$ is symmetric with respect to the principal axes $x$ and $y$. Figure \ref{fig11} shows that the electric field intensity is dominant in the outer vicinities of the top and bottom parts of the surfaces of the fibers, and is significant in the area between the fiber surfaces.

\begin{figure}[tbh]
	\begin{center}
		\includegraphics{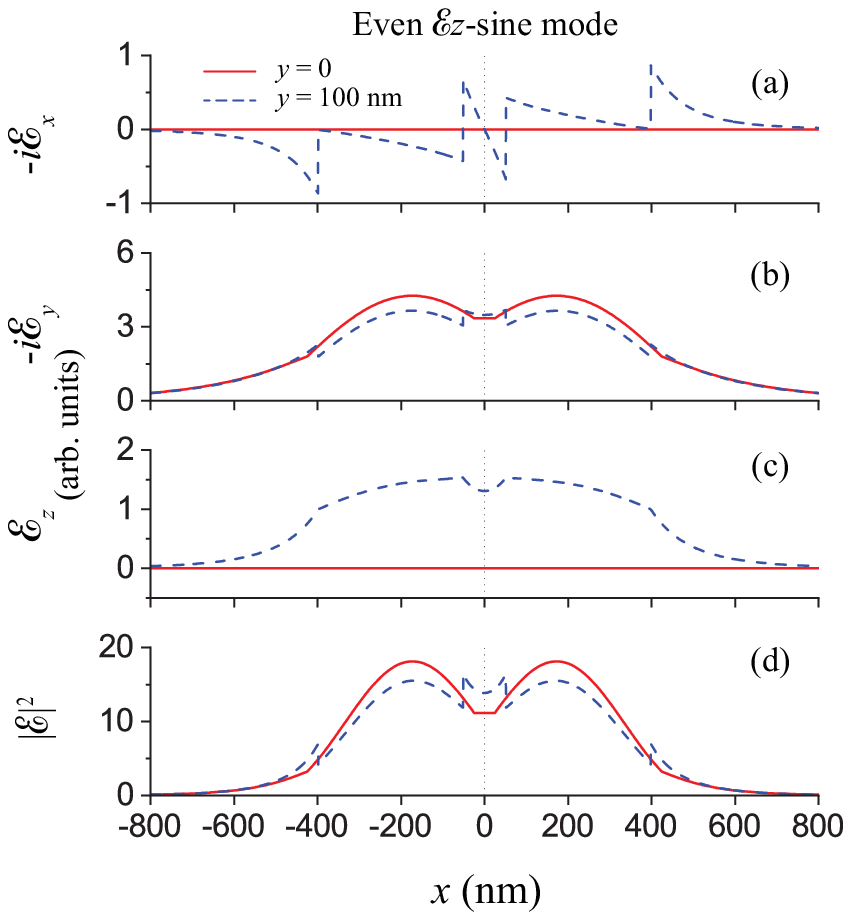}
	\end{center}
	\caption{(Color online) 
		Dependencies of the components $\mathcal{E}_x$, $\mathcal{E}_y$, and $\mathcal{E}_z$ and the intensity $|\mathcal{E}|^2$ of the electric field in the even $\mathcal{E}_z$-sine mode of two identical fibers on the $x$ coordinate. The fiber radius is $a=200$ nm. The fiber separation distance is $d=50$ nm. The parameters used are the same as for Fig.~\ref{fig11}(a).
	}
	\label{fig12}
\end{figure}

\begin{figure}[tbh]
	\begin{center}
		\includegraphics{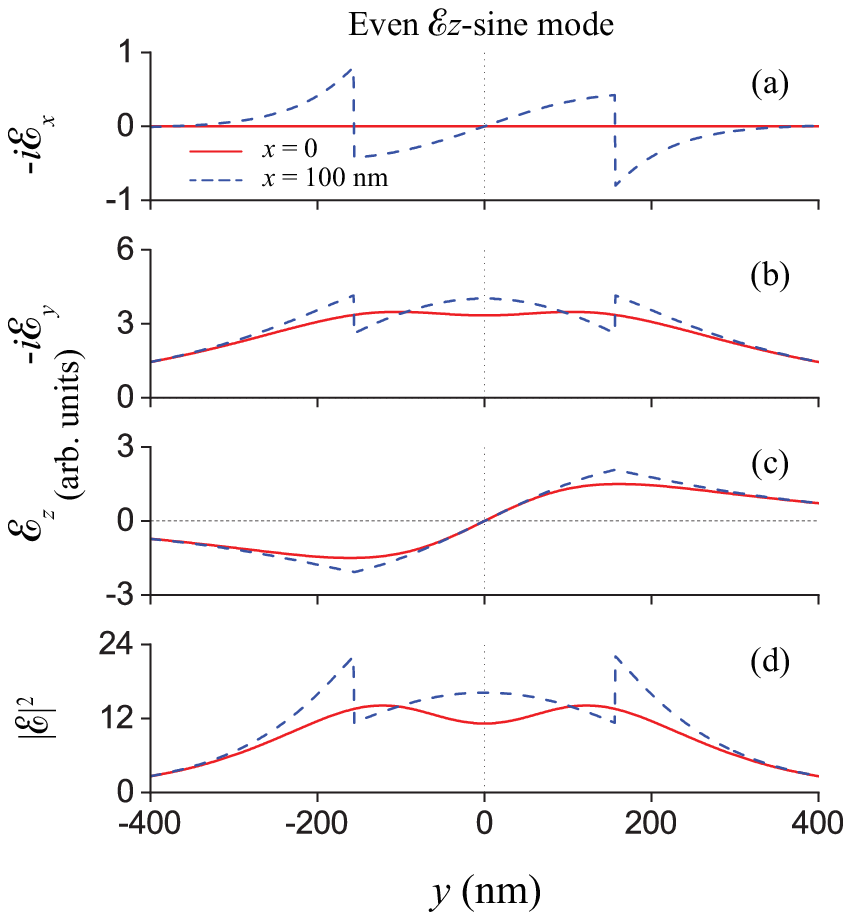}
	\end{center}
	\caption{(Color online) 
		Dependencies of the components $\mathcal{E}_x$, $\mathcal{E}_y$, and $\mathcal{E}_z$ and the intensity $|\mathcal{E}|^2$ of the electric field in the even $\mathcal{E}_z$-sine mode of two identical fibers on the $y$ coordinate. The fiber radius is $a=200$ nm. The fiber separation distance is $d=50$ nm. The parameters used are the same as for Fig.~\ref{fig11}(a).
	}
	\label{fig13}
\end{figure}

We plot in Figs.~\ref{fig12} and \ref{fig13} the dependencies of the components $\mathcal{E}_x$, $\mathcal{E}_y$, and $\mathcal{E}_z$ and the intensity $|\mathcal{E}|^2$ of the electric field in the even $\mathcal{E}_z$-sine mode on the $x$ and $y$ coordinates. Figure \ref{fig12} shows that $\mathcal{E}_x$ is antisymmetric and $\mathcal{E}_y$ and $\mathcal{E}_z$ are symmetric with respect to the coordinate $x$. We observe from Fig.~\ref{fig13} that $\mathcal{E}_x$ and $\mathcal{E}_z$ are antisymmetric and $\mathcal{E}_y$ is symmetric with respect to the coordinate $y$.
We see from the scales of the vertical axes in Figs.~\ref{fig12} and \ref{fig13} that 
all the three components $\mathcal{E}_x$, $\mathcal{E}_y$, and $\mathcal{E}_z$ of the field are significant, while
$\mathcal{E}_y$ [see Figs.~\ref{fig12}(b) and \ref{fig13}(b)] is dominant. These features are in agreement with the fact that, in the case of single fibers, the $\mathcal{E}_z$-sine modes are quasilinearly polarized along the $y$ axis \cite{Snyder1983,Marcuse1989,Okamoto2006}.

\subsubsection{Odd $\mathcal{E}_z$-sine mode}

\begin{figure}[tbh]
	\begin{center}
		\includegraphics{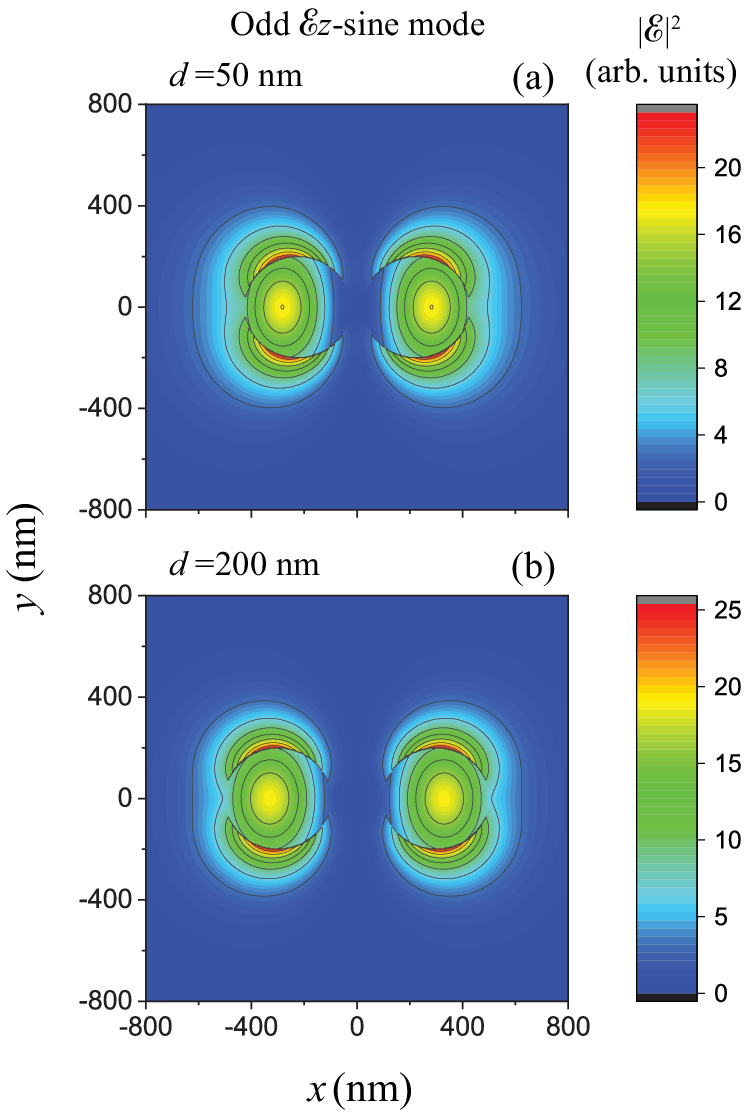}
	\end{center}
	\caption{(Color online) 
		Cross-sectional profile of the electric intensity distribution 
		$|\mathcal{E}|^2$ of the field in the odd $\mathcal{E}_z$-sine mode of two identical parallel nanofibers. The parameters used are the same as for Fig.~\ref{fig5}.
	}
	\label{fig14}
\end{figure}

We display in Fig.~\ref{fig14} the cross-sectional profile of the electric intensity distribution 
$|\mathcal{E}|^2$ of the field in the odd $\mathcal{E}_z$-sine mode of two identical parallel nanofibers. The figure shows that the intensity distribution $|\mathcal{E}|^2$ is symmetric with respect to the principal axes $x$ and $y$. We observe that the electric field intensity is dominant in the vicinities of the top and bottom parts of the surfaces of the fibers, and is significant in the outer vicinities of the left-side surface of the left-side fiber and the right-side surface of the right-side fiber.

\begin{figure}[tbh]
	\begin{center}
		\includegraphics{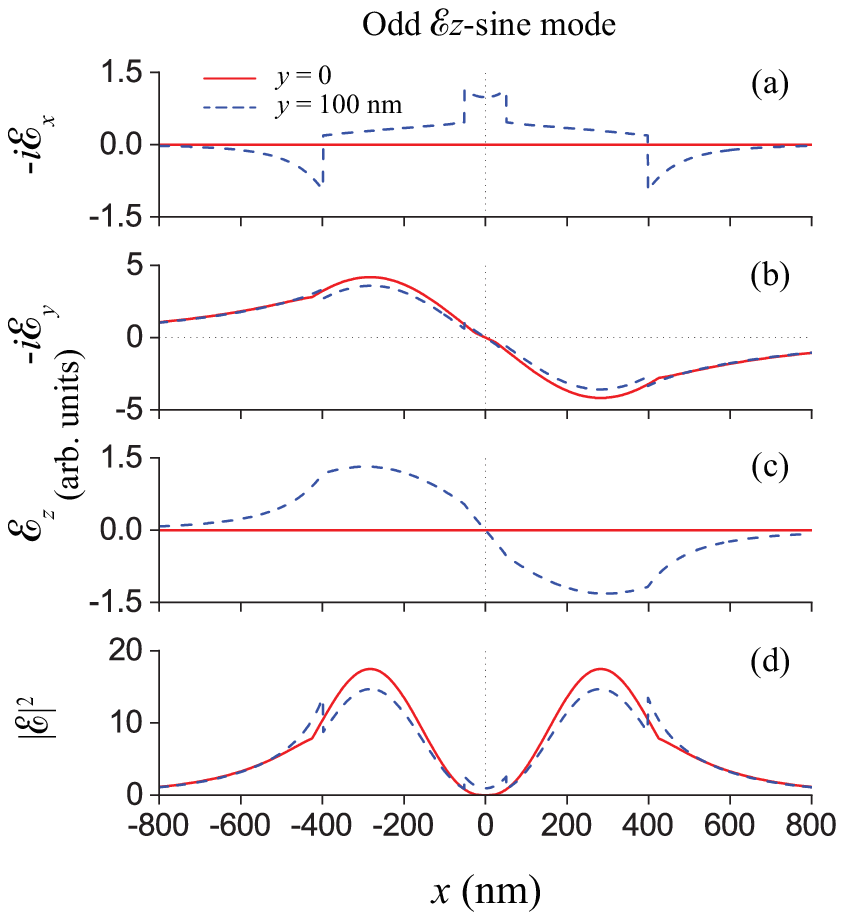}
	\end{center}
	\caption{(Color online) 
		Dependencies of the components $\mathcal{E}_x$, $\mathcal{E}_y$, and $\mathcal{E}_z$ and the intensity $|\mathcal{E}|^2$ of the electric field in the odd $\mathcal{E}_z$-sine mode of two identical fibers on the $x$ coordinate. The fiber radius is $a=200$ nm. The fiber separation distance is $d=50$ nm. The parameters used are the same as for Fig.~\ref{fig14}(a).
	}
	\label{fig15}
\end{figure}

\begin{figure}[tbh]
	\begin{center}
		\includegraphics{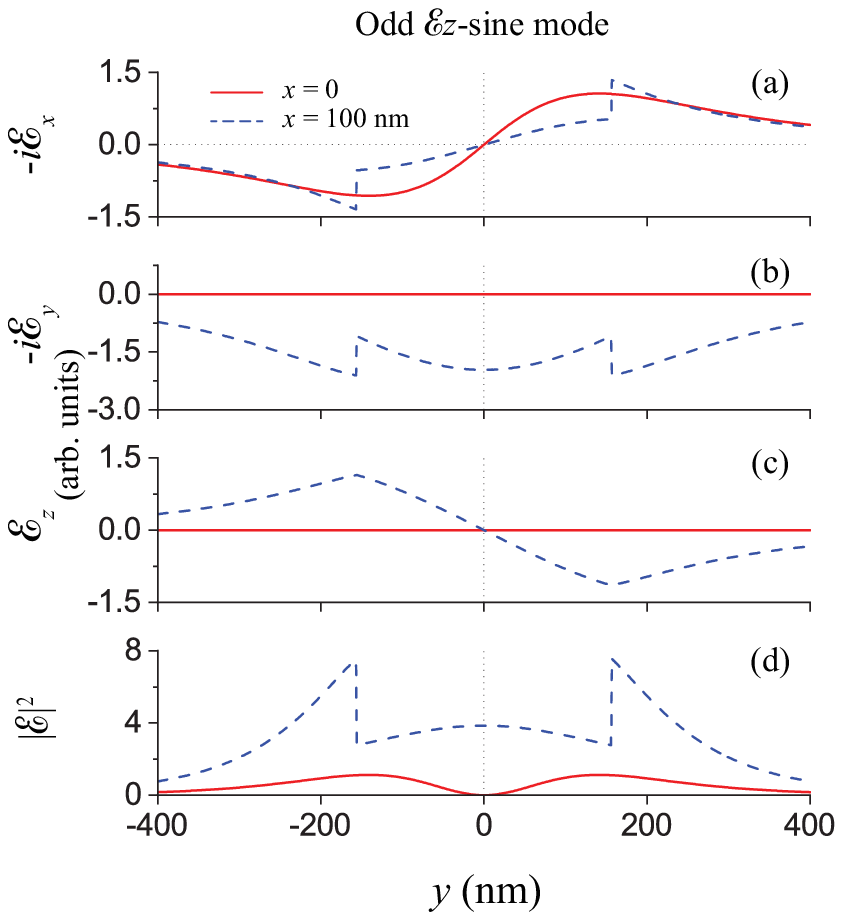}
	\end{center}
	\caption{(Color online) 
		Dependencies of the components $\mathcal{E}_x$, $\mathcal{E}_y$, and $\mathcal{E}_z$ and the intensity $|\mathcal{E}|^2$ of the electric field in the odd $\mathcal{E}_z$-sine mode of two identical fibers on the $y$ coordinate. The fiber radius is $a=200$ nm. The fiber separation distance is $d=50$ nm. The parameters used are the same as for Fig.~\ref{fig14}(a).
	}
	\label{fig16}
\end{figure}

We plot in Figs.~\ref{fig15} and \ref{fig16} the dependencies of the components $\mathcal{E}_x$, $\mathcal{E}_y$, and $\mathcal{E}_z$ and the intensity $|\mathcal{E}|^2$ of the electric field in the odd $\mathcal{E}_z$-sine mode on the $x$ and $y$ coordinates. Figure \ref{fig15} shows that $\mathcal{E}_x$ is symmetric and $\mathcal{E}_y$ and $\mathcal{E}_z$ are antisymmetric with respect to the coordinate $x$. We observe from Fig.~\ref{fig16} that $\mathcal{E}_x$ and $\mathcal{E}_z$ are antisymmetric and $\mathcal{E}_y$ is symmetric with respect to the coordinate $y$.
We see from the scales of the vertical axes in Figs.~\ref{fig15} and \ref{fig16} that all the three components $\mathcal{E}_x$, $\mathcal{E}_y$, and $\mathcal{E}_z$ of the field are significant.

Figure \ref{fig14} and the solid curves of Figs.~\ref{fig15}(d) and \ref{fig16}(d) confirm the prediction that the electric field of the odd $\mathcal{E}_z$-sine mode at the two-fiber center  $(x,y)=(0,0)$ is zero. This feature of the odd $\mathcal{E}_z$-sine mode can be used to trap ground-state atoms with a blue-detuned optical dipole potential \cite{Nobel prizers a,Nobel prizers b,Nobel prizers c}, or to trap Rydberg atoms with a ponderomotive optical Rydberg-electron potential \cite{ponderomotive 1,ponderomotive 2}. We emphasize that the existence of a local minimum of exact zero at the two-fiber center is a specific property of the intensity of the electric field in the odd $\mathcal{E}_z$-sine mode, and occurs for any fiber separation distance $d$.  To reduce to effects of the fiber surfaces on the atoms at the two-fiber center, we can increase the fiber separation distance.

We note that the symmetry properties of the field components, shown in Figs.~\ref{fig5}--\ref{fig16},
are in agreement with Eqs.~(\ref{g18})--(\ref{g21}). In particular, the symmetry properties of the curves in Fig.~\ref{fig6} (for the even $\mathcal{E}_z$-cosine mode) with respect to the coordinate $x$ are the same as those of the curves in Fig.~\ref{fig15} (for the odd $\mathcal{E}_z$-sine mode) and are opposite to those of the curves in Figs.~\ref{fig9} (for the odd $\mathcal{E}_z$-cosine mode) and \ref{fig12} (for the even $\mathcal{E}_z$-sine mode). Meanwhile, the symmetry properties of the curves in Fig.~\ref{fig7} (for the even $\mathcal{E}_z$-cosine mode) with respect to the coordinate $y$ are the same as those of the curves in Fig.~\ref{fig10} (for the odd $\mathcal{E}_z$-cosine mode) and are opposite to those of the curves in Figs.~\ref{fig13} (for the even $\mathcal{E}_z$-sine mode) and \ref{fig16} (for the odd $\mathcal{E}_z$-sine mode).

\subsection{Comparison between the exact mode theory and the coupled mode theory}

According to the coupled mode theory (CMT) \cite{Snyder1983,Marcuse1989,Okamoto2006},
the coupling length for the fiber modes with the principal polarization $p=x,y$ of two parallel fibers is $L_p=\pi/2\eta_p$, where $\eta_p=(\kappa_p-c_p\chi_p)/(1-c_p^2)$ is the power transfer coefficient. Here, $\kappa_p$, $\chi_p$, and $c_p$ are the coefficients of directional coupling, self coupling, and butt coupling, respectively.
The calculations of the power transfer coefficient $\eta_p$ for two parallel nanofibers have been reported in  \cite{CMT}.

In the framework of the coupled mode theory \cite{Snyder1983,Marcuse1989,Okamoto2006},
the transfer of power between the coupled fibers is a result of the beating between two normal modes. The power transfer coefficient is $\eta_p=(\beta_a^{(p)}-\beta_b^{(p)})/2$, where $\beta_a^{(p)}$ and $\beta_b^{(p)}$ are the propagation constants of the normal modes $a$ and $b$ with the principal polarization $p=x$ or $y$ (the $\mathcal{E}_z$-cosine or $\mathcal{E}_z$-sine type).
For two identical nanofibers, the normal modes $a$ and $b$ are the even ($\nu=-1$) and odd ($\nu=1$) modes, respectively. Hence, the power transfer coefficient $\eta_p$ can be calculated from the exact theory for the guided array modes \cite{Wijngaard1973}.

\begin{figure}[tbh]
	\begin{center}
		\includegraphics{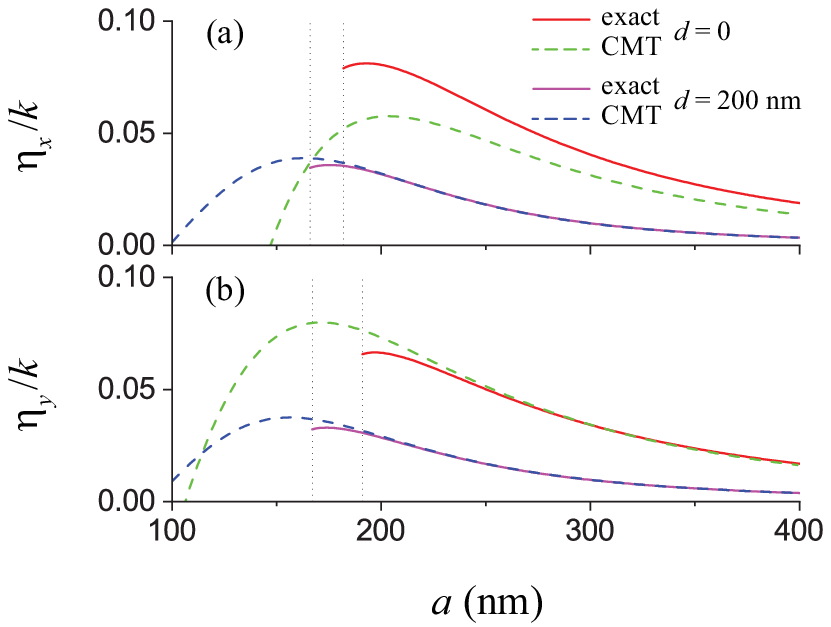}
	\end{center}
	\caption{(Color online) Power transfer coefficients $\eta_x$ (a) and $\eta_y$ (b) between two identical fibers calculated from the exact theory for the guided array modes (solid lines) and the coupled mode theory (dashed lines) as functions of the fiber radius $a$ for the fiber separation distances $d=0$ and 200 nm. The light wavelength is $\lambda=800$ nm. Other parameters are as in Fig.~\ref{fig2}. The vertical dashed lines indicate the positions of the cutoffs of the odd array modes. 
	}
	\label{fig17}
\end{figure}

\begin{figure}[tbh]
	\begin{center}
		\includegraphics{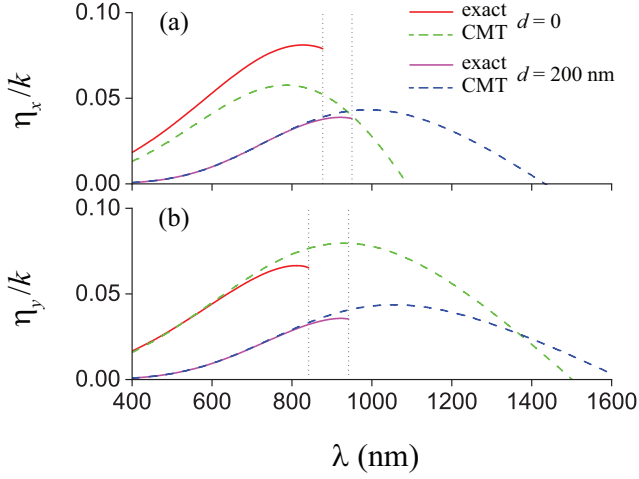}
	\end{center}
	\caption{(Color online) Power transfer coefficients $\eta_x$ (a) and $\eta_y$ (b) between two identical fibers calculated from the exact theory for the guided array modes (solid lines) and the coupled mode theory (dashed lines) as functions of the light wavelength $\lambda$ for the fiber separation distances $d=0$ and 200 nm. The fiber radius is $a=200$ nm. Other parameters are as in Fig.~\ref{fig3}. The vertical dashed lines indicate the positions of the cutoffs of the odd array modes. 
	}
	\label{fig18}
\end{figure}

\begin{figure}[tbh]
	\begin{center}
		\includegraphics{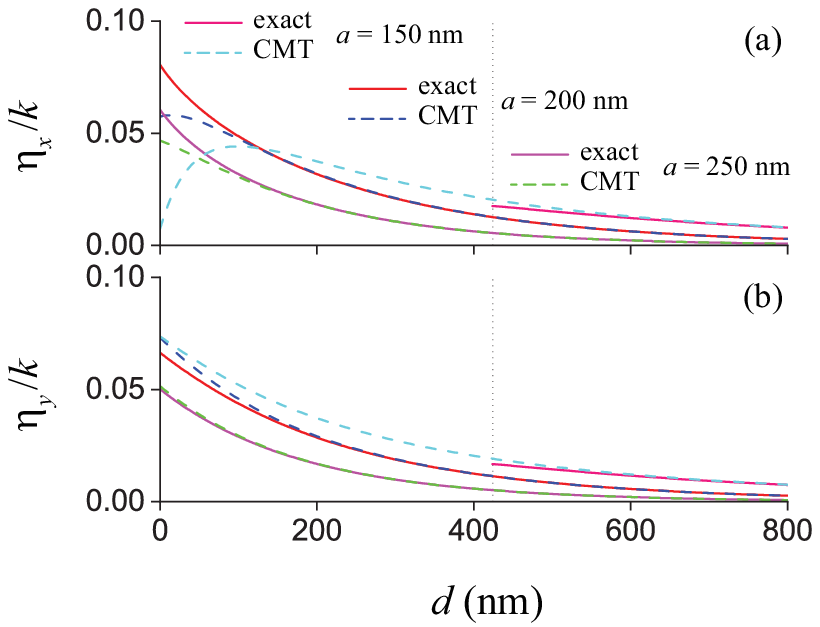}
	\end{center}
	\caption{(Color online) Power transfer coefficients $\eta_x$ (a) and $\eta_y$ (b) between two identical fibers calculated from the exact theory for the guided array modes (solid lines) and  the coupled mode theory (dashed lines) as functions of the fiber separation distance $d$ for the fiber radii $a=150$, 200, and 250 nm. The light wavelength is $\lambda=800$ nm. Other parameters are as in Fig.~\ref{fig2}. The vertical dashed lines indicate the positions of the cutoffs of the odd array modes. 
	}
	\label{fig19}
\end{figure} 

We plot in Figs.~\ref{fig17}--\ref{fig19} the power transfer coefficients $\eta_x$  and $\eta_y$ between two identical fibers calculated from the exact theory for the guided array modes (solid curves) and the coupled mode theory (dashed curves) as functions of the fiber radius $a$, the light wavelength $\lambda$, and the fiber separation distance $d$.
In these figures, the vertical dashed lines indicate the positions of the cutoffs of the odd array modes 
that occur in the framework of the exact theory when $d$ is small and either $a$ is small or $\lambda$ is large in the ranges plotted. The coupled mode theory is not able to predict the mode cutoffs.  Below a cutoff, the overlap between the modes of the individual fibers is so significant that only even array modes exist. In this regime, the concept of power transfer between the fibers becomes meaningless and, hence, the coupled mode theory is not valid. We observe from Figs.~\ref{fig17}--\ref{fig19} that the results of the coupled mode theory agree well with that of the exact theory in the far-above-cutoff regions where the fiber radius is large, the light wavelength is small, or the fiber separation distance $d$ is large. Near to a cutoff, the differences between the results of the exact and coupled mode theories
are significant but not dramatic. For touching nanofibers ($d=0$), the differences between the results of the two theories become large when the fiber radius $a$ is small (see Fig.~\ref{fig17}) or the light wavelength $\lambda$ is large (see Fig.~\ref{fig18}). Comparison between Figs.~\ref{fig17}(a) and \ref{fig17}(b) and between Figs.~\ref{fig18}(a) and \ref{fig18}(b) shows that, for touching nanofibers, the differences between the results of the two theories for the $x$-polarized modes are larger than those for
the $y$-polarized modes.

\subsection{Nonidentical nanofibers}

When the two nanofibers are not identical, the symmetry relations (\ref{g18}) and (\ref{g19}) for the $x$ coordinate are not valid and, hence, the guided array modes cannot be identified as even and odd modes anymore. We follow \cite{Wijngaard1973} and label the first two modes of the $\mathcal{E}_z$-sine type as $I_0^{(\mathrm{sin})}$ and $I_\pi^{(\mathrm{sin})}$. Similarly, we label the first two modes of the $\mathcal{E}_z$-cosine type as $I_0^{(\mathrm{cos})}$ and $I_\pi^{(\mathrm{cos})}$.
Like the case of identical nanofibers, modes $I_0^{(\mathrm{sin})}$ and $I_0^{(\mathrm{cos})}$  have no cutoff,
while modes $I_\pi^{(\mathrm{sin})}$ and $I_\pi^{(\mathrm{cos})}$ may have cutoffs.
Note that the symmetry relations (\ref{g20}) and (\ref{g21}) for the $y$ coordinate remain valid for nonidentical nanofibers. 

\begin{figure}[tbh]
	\begin{center}
		\includegraphics{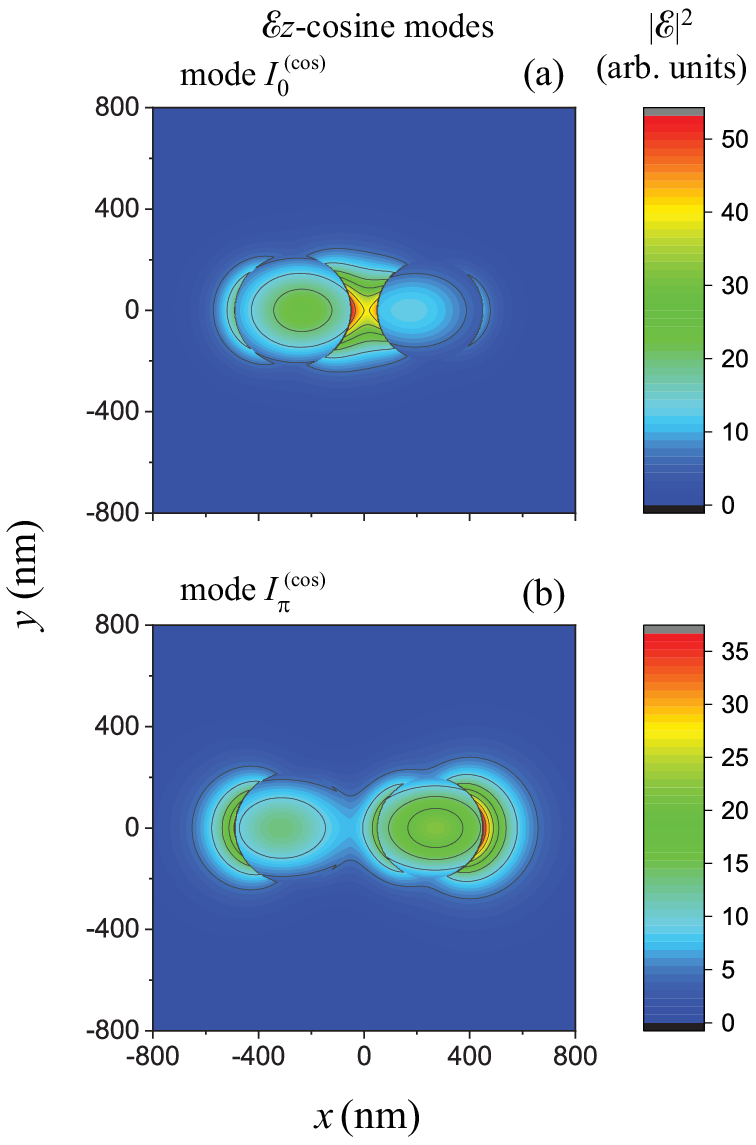}
	\end{center}
	\caption{(Color online) Cross-sectional profiles of the electric intensity distributions 
		$|\mathcal{E}|^2$ of the fields in the $\mathcal{E}_z$-cosine modes $I_0^{(\mathrm{cos})}$ and $I_{\pi}^{(\mathrm{cos})}$ of two nonidentical parallel nanofibers. The fiber radii are $a_1=220$ nm and $a_2=200$ nm. The fiber separation distance is $d=100$ nm. Other parameters are as in Fig.~\ref{fig2}.}
	\label{fig20}
\end{figure}

\begin{figure}[tbh]
	\begin{center}
		\includegraphics{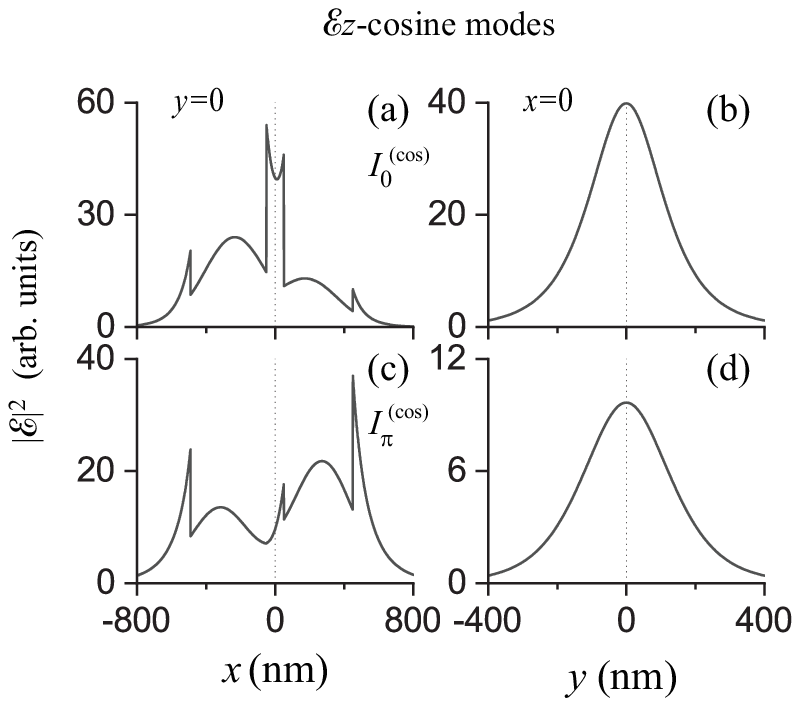}
	\end{center}
	\caption{ 
		Dependencies of the intensities $|\mathcal{E}|^2$ of the electric fields in the  $\mathcal{E}_z$-cosine modes $I_0^{(\mathrm{cos})}$ (upper row) and $I_{\pi}^{(\mathrm{cos})}$ (lower row) of two nonidentical parallel nanofibers on the $x$ (left column) and $y$ (right column) coordinates. The parameters used are the same as for Fig.~\ref{fig20}.
	}
	\label{fig21}
\end{figure}

\begin{figure}[tbh]
	\begin{center}
		\includegraphics{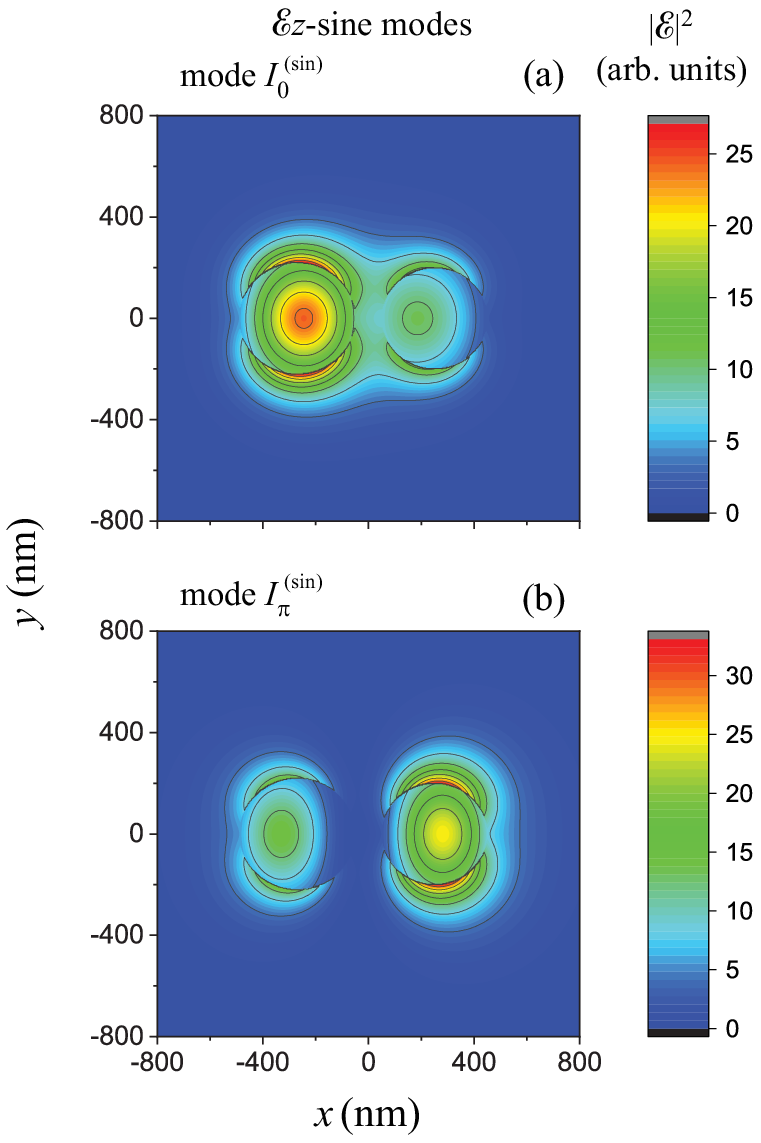}
	\end{center}
	\caption{(Color online) Cross-sectional profiles of the electric intensity distributions 
		$|\mathcal{E}|^2$ of the fields in the $\mathcal{E}_z$-sine modes $I_0^{(\mathrm{sin})}$ and $I_{\pi}^{(\mathrm{sin})}$ of two nonidentical parallel nanofibers. The parameters used are the same as for Fig.~\ref{fig20}.}
	\label{fig22}
\end{figure}

\begin{figure}[tbh]
	\begin{center}
		\includegraphics{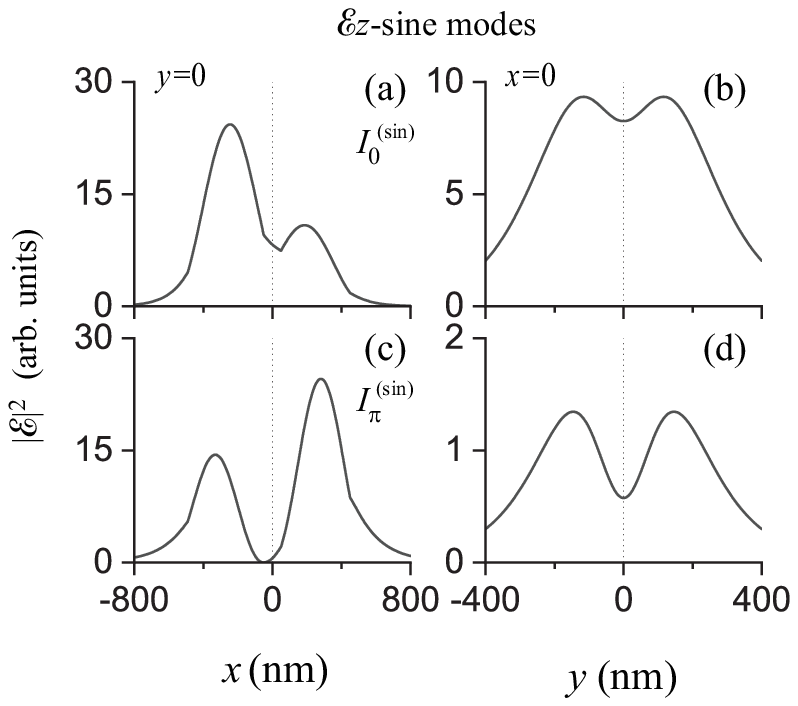}
	\end{center}
	\caption{ 
		Dependencies of the intensities $|\mathcal{E}|^2$ of the electric fields in the  $\mathcal{E}_z$-sine modes $I_0^{(\mathrm{sin})}$ (upper row) and $I_{\pi}^{(\mathrm{sin})}$ (lower row) of two nonidentical parallel nanofibers on the $x$ (left column) and $y$ (right column) coordinates. The parameters used are the same as for Fig.~\ref{fig22}.
	}
	\label{fig23}
\end{figure}

We plot in Fig.~\ref{fig20} the cross-sectional profiles of the electric intensity distributions $|\mathcal{E}|^2$ of the fields in the $\mathcal{E}_z$-cosine modes of two nonidentical parallel nanofibers.
We show in Fig.~\ref{fig22} the dependencies of $|\mathcal{E}|^2$ on the $x$ and $y$ coordinates. The corresponding results for the $\mathcal{E}_z$-sine modes
are shown in Figs.~\ref{fig22} and \ref{fig23}.
We observe from the figures that the intensity distribution $|\mathcal{E}|^2$ is asymmetric with respect to the coordinate $x$ and symmetric with respect to the coordinate $y$. Figure \ref{fig20}(a) shows that the electric field intensity is dominant in the area between the nanofibers and, hence, atoms can be attracted to this area using a single red-detuned light field.  
Figures \ref{fig20}--\ref{fig23} show that, in the cases of $I_0^{(\mathrm{cos,sin})}$ modes, the intensity of the field  in the area of the bigger nanofiber (nanofiber 1) is higher than that in the area of the smaller nanofiber (nanofiber 2).  Meanwhile, in the cases of $I_\pi^{(\mathrm{cos,sin})}$ modes, the intensity of the field in the area of the smaller nanofiber (nanofiber 2) is higher than that in the area of the bigger nanofiber (nanofiber 1).

\section{Summary}
\label{sec:summary}

We have studied the cross-sectional profiles and spatial distributions of the fields in guided normal modes of two coupled parallel optical nanofibers. We have shown that the distributions of the components of the field in a guided normal mode of two identical nanofibers are either symmetric or antisymmetric with respect to the radial principal axis $x$ and the tangential principal axis $y$ in the  cross-sectional plane of the fibers. The symmetry of the magnetic field components  with respect to the principal axes is opposite to that of the electric field components. We have found that, in the case of even $\mathcal{E}_z$-cosine modes, the electric intensity distribution is dominant in the area between the fibers, with a saddle point at the two-fiber center. This feature may be used to attract atoms with a single red-detuned guided normal-mode light field. Meanwhile, in the case of odd $\mathcal{E}_z$-sine modes, the electric intensity distribution at the two-fiber center attains a local minimum of exactly zero. This feature may be used to trap atoms with a single blue-detuned guided normal-mode light field. We have observed that the differences between the results of the coupled mode theory and the exact theory are large when the fiber separation distance is small and either the fiber radius is small or the light wavelength is large. We have shown that, in the case where the two nanofibers are not identical, the intensity distribution is symmetric about the radial principal axis $x$ and asymmetric about the tangential principal axis $y$. Potential uses of the fields in guided normal modes of two coupled parallel nanofibers for trapping, guiding, and probing atoms deserve further systematic investigations.

\begin{acknowledgments}
This work was supported by the Okinawa Institute of Science and Technology Graduate University.
\end{acknowledgments}


\appendix

\section{Equations for the expansion coefficients of the array modes}
\label{appendix}

For the $\mathcal{E}_z$-cosine modes, the expansion coefficients $E_{nj}$, $F_{nj}$, $G_{nj}$, and $H_{nj}$ vanish. For these modes, the coefficients $A_{nj}$ and $B_{nj}$ for the field inside the fibers are given by the equations
\begin{eqnarray}\label{g5}
	J_n(u_j)A_{nj}&=&K_n(w_j)C_{nj}+I_n(w_j)\sum_{m=0}^\infty f_{jnm}C_{m\bar{j}},\nonumber\\
	J_n(u_j)B_{nj}&=&K_n(w_j)D_{nj}+I_n(w_j)\sum_{m=0}^\infty g_{jnm}D_{m\bar{j}},\qquad
\end{eqnarray}	
while the coefficients $C_{nj}$ and $D_{nj}$ for the field outside the fibers are nonzero solutions of the equations \cite{Wijngaard1973}
\begin{eqnarray}\label{g6}
	&&	n\bigg(\frac{1}{u_j^2}+\frac{1}{w_j^2}\bigg)\bigg[ K_n(w_j)C_{nj}	
	+I_n(w_j)\sum_{m=0}^\infty f_{jnm}C_{m\bar{j}}\bigg]
	\nonumber\\&&\mbox{} 
	+\frac{\omega\mu_0}{\beta}\bigg[\frac{J'_n(u_j)}{u_jJ_n(u_j)}
	+\frac{K'_n(w_j)}{w_jK_n(w_j)}\bigg] K_n(w_j)D_{nj}
	\nonumber\\&&\mbox{}
	+\frac{\omega\mu_0}{\beta}\bigg[\frac{J'_n(u_j)}{u_jJ_n(u_j)}	
	+\frac{I'_n(w_j)}{w_jI_n(w_j)}\bigg]I_n(w_j) \sum_{m=0}^\infty g_{jnm}D_{m\bar{j}}
	\nonumber\\&&
	=0,
	\nonumber\\	
	&&  n\bigg(\frac{1}{u_j^2}+\frac{1}{w_j^2}\bigg)
	\bigg[K_n(w_j)D_{nj}+I_n(w_j)\sum_{m=0}^\infty g_{jnm}D_{m\bar{j}}\bigg]
	\nonumber\\&&\mbox{}
	+\frac{\omega\epsilon_0}{\beta}\bigg[\frac{n_j^2J'_n(u_j)}{u_jJ_n(u_j)}
	+\frac{n_0^2K'_n(w_j)}{w_jK_n(w_j)}\bigg]K_n(w_j)C_{nj}
	\nonumber\\&&\mbox{}
	+\frac{\omega\epsilon_0}{\beta}\bigg[\frac{n_j^2J'_n(u_j)}{ u_jJ_n(u_j)}
	+\frac{n_0^2I'_n(w_j)}{w_jI_n(w_j)}\bigg]
	I_n(w_j)\sum_{m=0}^\infty f_{jnm}C_{m\bar{j}}
	\nonumber\\&&
	=0.
\end{eqnarray}
Here, we have introduced the notation $\bar{j}=2$ or 1 for $j=1$ or 2, respectively.
We have also introduced the parameters
\begin{equation}\label{g7}
	u_j=h_ja_j,\qquad w_j=qa_j, 
\end{equation}
and the coefficients
\begin{equation}\label{g8}
	f_{1nm}=(-1)^m f_{nm},\qquad
	g_{1nm}=(-1)^m g_{nm},
\end{equation}
and
\begin{equation}\label{g9}
	f_{2nm}=(-1)^{n}f_{nm},\qquad 
	g_{2nm}=(-1)^{n} g_{nm},
\end{equation}
where
\begin{eqnarray}\label{g10}
	f_{nm}&=& K_{m+n}(qW)+ K_{m-n}(qW) \mbox{ for } n>0,\nonumber\\
	f_{0,m}&=& K_{m}(qW),\nonumber\\
	g_{nm}&=& -K_{m+n}(qW)+ K_{m-n}(qW),
\end{eqnarray}
with $W=d+a_1+a_2$ being the distance between the fiber centers.

For the $\mathcal{E}_z$-sine modes, the expansion coefficients $A_{nj}$, $B_{nj}$, $C_{nj}$, and $D_{nj}$ vanish. For these modes, the coefficients $E_{nj}$ and $F_{nj}$ for the field inside the fibers are given by the equations
\begin{eqnarray}\label{g11}
	J_n(u_j)E_{nj}&=&K_n(w_j)G_{nj}+I_n(w_j)\sum_{m=0}^\infty g_{jnm}G_{m\bar{j}},\nonumber\\
	J_n(u_j)F_{nj}&=&K_n(w_j)H_{nj}+I_n(w_j)\sum_{m=0}^\infty f_{jnm}H_{m\bar{j}},\qquad
\end{eqnarray} 	
while the coefficients $G_{nj}$ and $H_{nj}$ for the field outside the fibers are nonzero solutions of the equations \cite{Wijngaard1973}
\begin{eqnarray}\label{g12}
	&&	n\bigg(\frac{1}{u_j^2}+\frac{1}{w_j^2}\bigg)\bigg[ K_n(w_j)G_{nj}+I_n(w_j)\sum_{m=0}^\infty g_{jnm}G_{m\bar{j}}\bigg]
	\nonumber\\&&\mbox{}  
	-\frac{\omega\mu_0}{\beta}\bigg[\frac{J'_n(u_j)}{u_jJ_n(u_j)}
	+\frac{K'_n(w_j)}{w_jK_n(w_j)}\bigg] K_n(w_j)H_{nj}
	\nonumber\\&&\mbox{}	 
	-\frac{\omega\mu_0}{\beta }\bigg[\frac{J'_n(u_j)}{u_jJ_n(u_j)}
	+\frac{I'_n(w_j)}{w_jI_n(w_j)}\bigg]I_n(w_j)\sum_{m=0}^\infty f_{jnm}H_{m\bar{j}}
	\nonumber\\&&
	=0,
	\nonumber\\
	&&  n\bigg(\frac{1}{u_j^2}+\frac{1}{w_j^2}\bigg)\bigg[K_n(w_j)H_{nj}
	+I_n(w_j)\sum_{m=0}^\infty f_{jnm}H_{m\bar{j}}\bigg]
	\nonumber\\&&\mbox{}
	-\frac{\omega\epsilon_0}{\beta}\bigg[\frac{n_j^2J'_n(u_j)}{u_jJ_n(u_j)}
	+\frac{n_0^2K'_n(w_j)}{w_jK_n(w_j)}\bigg]K_n(w_j)G_{nj}
	\nonumber\\&&\mbox{}
	-\frac{\omega\epsilon_0}{\beta}\bigg[\frac{n_j^2J'_n(u_j)}{u_jJ_n(u_j)}
	+\frac{n_0^2I'_n(w_j)}{w_jI_n(w_j)}\bigg]
	I_n(w_j)\sum_{m=0}^\infty g_{jnm}G_{m\bar{j}}
	\nonumber\\&&
	=0.
\end{eqnarray}	

We consider the particular case where the two fibers are identical, that is, the two fibers have the same radius $a_1=a_2$ and the same core refractive index $n_1=n_2$. In this case, we have $h_1=h_2$, $u_1=u_2$, and $w_1=w_2$.
Then, for the $\mathcal{E}_z$-cosine modes, we find
\begin{eqnarray}\label{g14a}
	A_{m2}&=& (-1)^m \nu A_{m1}, \qquad B_{m2}=(-1)^m \nu B_{m1},\nonumber\\ 
	C_{m2}&=& (-1)^m \nu C_{m1}, \qquad D_{m2}=(-1)^m \nu D_{m1}, \qquad
\end{eqnarray}  
and, for the $\mathcal{E}_z$-sine modes, we get
\begin{eqnarray}\label{g15a}
	E_{m2}&=& (-1)^m \nu E_{m1}, \qquad F_{m2}=(-1)^m \nu F_{m1},\nonumber\\
	G_{m2}&=& (-1)^m \nu G_{m1}, \qquad H_{m2}=(-1)^m \nu H_{m1}, \qquad
\end{eqnarray}  
where $\nu=-1$ or $+1$ corresponds to the even or odd mode \cite{Wijngaard1973}. 
Hence, Eqs.~(\ref{g6}) for the $\mathcal{E}_z$-cosine modes take the form \cite{Wijngaard1973}
\begin{eqnarray}\label{g16}
	&&	n\bigg(\frac{1}{u^2}+\frac{1}{w^2}\bigg)\bigg[ K_n(w)C_{n1}	
	+I_n(w)\sum_{m=0}^\infty \nu f_{nm}C_{m1}\bigg]
	\nonumber\\&&\mbox{} 
	+\frac{\omega\mu_0}{\beta}\bigg[\frac{J'_n(u)}{uJ_n(u)}
	+\frac{K'_n(w)}{wK_n(w)}\bigg] K_n(w)D_{n1}
	\nonumber\\&&\mbox{}
	+\frac{\omega\mu_0}{\beta}\bigg[\frac{J'_n(u)}{uJ_n(u)}	
	+\frac{I'_n(w)}{wI_n(w)}\bigg]I_n(w) \sum_{m=0}^\infty \nu g_{nm}D_{m1}
	=0,
	\nonumber\\	
	&&  n\bigg(\frac{1}{u^2}+\frac{1}{w^2}\bigg)
	\bigg[K_n(w)D_{n1}+I_n(w)\sum_{m=0}^\infty \nu g_{nm}D_{m1}\bigg]
	\nonumber\\&&\mbox{}
	+\frac{\omega\epsilon_0}{\beta}\bigg[\frac{n_1^2J'_n(u)}{uJ_n(u)}
	+\frac{n_0^2K'_n(w)}{wK_n(w)}\bigg]K_n(w)C_{n1}
	\nonumber\\&&\mbox{}
	+\frac{\omega\epsilon_0}{\beta}\bigg[\frac{n_1^2J'_n(u)}{ uJ_n(u)}
	+\frac{n_0^2I'_n(w)}{wI_n(w)}\bigg]
	I_n(w)\sum_{m=0}^\infty \nu f_{nm}C_{m1}
	=0,
	\nonumber\\
\end{eqnarray}
and Eqs.~(\ref{g12}) for the $\mathcal{E}_z$-sine modes become \cite{Wijngaard1973} 
\begin{eqnarray}\label{g17}
	&&	n\bigg(\frac{1}{u^2}+\frac{1}{w^2}\bigg)\bigg[ K_n(w)G_{n1}+I_n(w)\sum_{m=0}^\infty \nu g_{nm}G_{m1}\bigg]
	\nonumber\\&&\mbox{}  
	-\frac{\omega\mu_0}{\beta}\bigg[\frac{J'_n(u)}{uJ_n(u)}
	+\frac{K'_n(w)}{wK_n(w)}\bigg] K_n(w)H_{n1}
	\nonumber\\&&\mbox{}	 
	-\frac{\omega\mu_0}{\beta }\bigg[\frac{J'_n(u)}{uJ_n(u)}
	+\frac{I'_n(w)}{wI_n(w)}\bigg]I_n(w)\sum_{m=0}^\infty \nu f_{nm}H_{m1}
	=0,
	\nonumber\\
	&&  n\bigg(\frac{1}{u^2}+\frac{1}{w^2}\bigg)\bigg[K_n(w)H_{n1}
	+I_n(w)\sum_{m=0}^\infty \nu f_{nm}H_{m1}\bigg]
	\nonumber\\&&\mbox{}
	-\frac{\omega\epsilon_0}{\beta}\bigg[\frac{n_1^2J'_n(u)}{uJ_n(u)}
	+\frac{n_0^2K'_n(w)}{wK_n(w)}\bigg]K_n(w)G_{n1}
	\nonumber\\&&\mbox{}
	-\frac{\omega\epsilon_0}{\beta}\bigg[\frac{n_1^2J'_n(u)}{uJ_n(u)}
	+\frac{n_0^2I'_n(w)}{wI_n(w)}\bigg]
	I_n(w)\sum_{m=0}^\infty \nu g_{nm}G_{m1}
	=0.
	\nonumber\\
\end{eqnarray}


\begin{thebibliography}{99}
	
\bibitem{Snyder1983} A. W. Snyder and J. D. Love, \textit{Optical Waveguide Theory} (Chapman and Hall, New York, 1983).

\bibitem{Marcuse1989} D. Marcuse, \textit{Light Transmission Optics} (Krieger, Malabar, 1989).

\bibitem{Okamoto2006} K. Okamoto, \textit{Fundamentals of Optical Waveguides} (Elsevier, New York, 2006).

\bibitem{TongNat03} L. Tong, R. R. Gattass, J. B. Ashcom, S. He, J. Lou, M. Shen, I. Maxwell, and E. Mazur, Nature (London) \textbf{426}, 816 (2003).

\bibitem{review2016} T. Nieddu, V. Gokhroo, and S. Nic Chormaic, J. Opt. \textbf{18}, 053001 (2016).

\bibitem{review2017} P. Solano, J. A. Grover, J. E. Homan, S. Ravets, F. K. Fatemi, L. A. Orozco, and S. L. Rolston, Adv. At. Mol. Opt. Phys. \textbf{66}, 439 (2017).

\bibitem{Nayak2018} K. Nayak, M. Sadgrove, R. Yalla, Fam Le Kien, and K. Hakuta, J. Opt. \textbf{20}, 073001 (2018).

\bibitem{onecolor}
V. I. Balykin, K. Hakuta, Fam Le Kien, J. Q. Liang, and  M. Morinaga,  Phys. Rev. A \textbf{70}, 011401(R) (2004); 

\bibitem{twocolor}
Fam Le Kien, V. I. Balykin, and K. Hakuta, Phys. Rev. A \textbf{70}, 063403 (2004).

\bibitem{Vetsch2010} E. Vetsch, D. Reitz, G. Sagu\'{e}, R. Schmidt, S. T. Dawkins, and A. Rauschenbeutel, Phys. Rev. Lett. \textbf{104}, 203603 (2010).

\bibitem{Goban2012} A. Goban, K. S. Choi, D. J. Alton, D. Ding, C. Lacro\^{u}te, M. Pototschnig, T. Thiele, N. P. Stern, and H. J. Kimble, Phys. Rev. Lett. \textbf{109}, 033603 (2012).

\bibitem{cesium decay} Fam Le Kien, S. Dutta Gupta, V. I. Balykin, and K. Hakuta, 
Phys. Rev. A \textbf{72}, 032509 (2005).

\bibitem{Nayak2007} K. P. Nayak, P. N. Melentiev, M. Morinaga, Fam Le Kien, V. I. Balykin, and K. Hakuta, Opt. Express \textbf{15}, 5431 (2007). 

\bibitem{Nayak2008}  K. P. Nayak and K. Hakuta,  New J. Phys. \textbf{10}, 053003 (2008).

\bibitem{absorption} Fam Le Kien, V. I. Balykin, and K. Hakuta, Phys. Rev. A \textbf{73}, 013819 (2006).

\bibitem{Sague2007} G. Sague, E. Vetsch, W. Alt, D. Meschede, and
A. Rauschenbeutel,  Phys. Rev. Lett. \textbf{99}, 163602 (2007).

\bibitem{Rajasree2020}  K. S. Rajasree, T. Ray, K. Karlsson, J. L. Everett, and S. Nic Chormaic, 
Phys. Rev. Res. \textbf{2}, 012038 (2020).

\bibitem{quadrupole} Fam Le Kien, T. Ray, T. Nieddu, T. Busch, and S. Nic Chormaic,  Phys. Rev. A \textbf{97}, 013821 (2018).

\bibitem{Ray2020} T. Ray, R. K. Gupta, V. Gokhroo, J. L. Everett, T. Nieddu, K. S. Rajasree, and S.  Nic Chormaic, New J. Phys. \textbf{22}, 062001 (2020).

\bibitem{slot} M. Daly, V. G. Truong, C. F. Phelan, K. Deasy, and S. Nic Chormaic, New J. Phys. \textbf{16}, 053052 (2014). 

\bibitem{Glorieux2019} C. Ding, V. Loo, S. Pigeon, R. Gautier, M. Joos, E. Wu,
E. Giacobino, A. Bramati, and Q. Glorieux, New J. Phys. \textbf{21}, 073060 (2019).

\bibitem{CMT} Fam Le Kien, L. Ruks, S. Nic Chormaic, and T. Busch,  New J. Phys. \textbf{22}, 123007 (2020).

\bibitem{Wijngaard1973} W. Wijngaard, J. Opt. Soc. Am.  \textbf{63}, 944 (1973).

\bibitem{Yamashita1985} E. Yamashita, S. Ozeki, and K. Atsuki,  J. Lightwave Technol. \textbf{3}, 341  (1985).

\bibitem{Kishi1989} N. Kishi, E. Yamashita, and H. Kawabata, J. Lightwave Technol. \textbf{7}, 902 (1989).

\bibitem{Huang1990} H. S. Huang and H. C. Chang, J. Lightwave Technol. \textbf{8},  945 (1990).

\bibitem{Chang1997a} C. S. Chang and H. C. Chang, J. Lightwave Technol. \textbf{15}, 1213 (1997).

\bibitem{Huang1989} H. S. Huang and H. C. Chang, Opt. Lett. \textbf{14}, 90 (1989).

\bibitem{Chang1997b} C. S. Chang and H. C. Chang, J. Lightwave Technol. \textbf{15}, 1225 (1997).

\bibitem{Lodahl2017} P. Lodahl, S. Mahmoodian, S. Stobbe, P. Schneeweiss,
J. Volz, A. Rauschenbeutel, H. Pichler, and P. Zoller, Nature \textbf{541}, 473 (2017).
	
\bibitem{Nobel prizers a} S. Chu, Rev. Mod. Phys. \textbf{70}, 685 (1998). 

\bibitem{Nobel prizers b} C. Cohen-Tannoudji, Rev. Mod. Phys. \textbf{70}, 707 (1998). 
	
\bibitem{Nobel prizers c} W. D. Phillips, Rev. Mod. Phys. \textbf{70}, 721 (1998).	

\bibitem{ponderomotive 1} S. K. Dutta, J. R. Guest, D. Feldbaum, A. Walz-Flannigan, and G. Raithel, Phys. Rev. Lett. \textbf{85}, 5551 (2000).

\bibitem{ponderomotive 2} D. Barredo, V. Lienhard, P. Scholl, S. de L\'{e}s\'{e}leuc, T. Boulier, A. Browaeys, and T. Lahaye, Phys. Rev. Lett. \textbf{124}, 023201 (2020).

\bibitem{Malitson} I. H. Malitson, J. Opt. Soc. Am. \textbf{55}, 1205 (1965).

\bibitem{Ghosh} G. Ghosh, Handbook of Thermo-Optic Coefficients of Optical
Materials with Applications (Academic Press, New York, 1997).

\end{thebibliography}
\end{document}